\documentclass[aps,pre,preprint,groupedaddress]{revtex4-1}

\usepackage{graphicx}
\usepackage{dsfont}
\usepackage{amssymb}
\usepackage{amsmath}
\usepackage{amsthm}
\usepackage{psfrag}
\usepackage{bm}

\begin{document}


\title{Phase transition in the Jarzynski estimator \\ of free energy differences}


\author{Alberto Su\'arez}
\altaffiliation{Permanent address: 
Computer Science Department, Escuela Polit\'ecnica Superior. Universidad Aut\'onoma de Madrid. Calle Francisco Tom\'as y Valiente, 11. 28049 Madrid (Spain)}
\email{alberto.suarez@uam.es}
\author{Robert Silbey}
\author{Irwin Oppenheim}
\email{irwin@mit.edu}
\affiliation{Chemistry Department \\ Massachusetts Institute of Technology \\ 77 Massachusetts Avenue \\
Cambridge, MA 02139 (USA)}


\date{\today}

\begin{abstract}
The transition between a regime in which thermodynamic relations 
apply only to ensembles of small systems coupled
to a large environment and a regime in which they can be used  
to characterize individual macroscopic systems is analyzed in terms of the 
change in behavior of the Jarzynski estimator of equilibrium 
free energy differences from nonequilibrium work measurements.
Given a fixed number of measurements, the Jarzynski estimator 
is unbiased for sufficiently small systems. In these 
systems the directionality of time is poorly defined and 
the configurations that dominate the empirical average, but 
which are in fact typical of the {\it reverse} process, are 
sufficiently well sampled. As the system size increases the arrow of time 
becomes better defined. The dominant atypical fluctuations become rare 
and eventually cannot be sampled with the limited 
resources that are available. Asymptotically, only typical work values
are measured. The Jarzynski estimator becomes maximally 
biased and approaches the exponential of minus the average work,
which is the result that is expected from standard macroscopic 
thermodynamics. In the proper scaling limit, this regime change 
has been recently described in terms of a phase transition in  
variants of the random energy model (REM). In this paper,
this correspondence is further demonstrated in two examples  
of physical interest: the sudden compression 
of an ideal gas and adiabatic quasi-static volume changes in a 
dilute real gas.  
\end{abstract}



\maketitle

\section{Introduction \label{sec:introduction}} 

The laws of thermodynamics summarize empirical observations about the
approximate and most probable behavior of macroscopic 
systems \cite{gibbs_1902_elementary,khinchin_1949_mathematical,zwanzig_2001_nonequilibrium,castiglione++_2008_chaos}.
They are formulated under the assumption of limited resources in
the measurements of the quantities involved. 
In the words of Gibbs, thermodynamics laws
``express the laws of mechanics for [systems 
of a great number of particles] as 
they appear to beings who have not the fineness of perception to enable 
them to appreciate quantities of the order of magnitude of those 
which relate to single particles, and who cannot repeat their experiments 
often enough to obtain any but the most probable results" 
\cite{gibbs_1902_elementary}. This statement summarizes
the conditions under which a proper thermodynamic description
for a single system can be made: 
(i) the system has a large number of degrees of freedom, 
(ii) the quantities of interest involve
averages over space and time on 
scales that are large compared to the corresponding 
molecular scales,
(iii) there are limitations in the time span of the measurements,
and
(iv) there are limitations in the number of measurements made. 
For instance, the elimination of condition (iii) gives rise
to the objection formulated by Zermelo, based on the 
fact that any isolated mechanical system will come 
arbitrarily close to its initial state provided that
a sufficiently long period of time elapses
(Poincar\'e recurrences)\cite{zermelo_1896_ueber,zwanzig_2001_nonequilibrium,castiglione++_2008_chaos}. 
If the system is ergodic,
averages over phase-space can be replaced by time averages 
and conditions (iii) and (iv) are equivalent. 

Improvements in measurement devices and techniques, 
an increasing interest in smaller scale systems in biology, 
physics and chemistry 
\cite{liphardt++_2002_equilibrium,barkai++_2004_theory,bustamante++_2005_nonequilibrium,ruschhaupt++_2006_one,thorn++_2008_experimental,ritort_2008_nonequilibrium} 
and developments in dynamical systems and
chaos theory \cite{gaspard_1998_chaos} 
have prompted numerous researchers to develop  
interpretations and  extensions of thermodynamics applicable
to systems with small numbers of degrees of freedom
\cite{oppenheim+mazur_1957_density,mazur+oppenheim_1957_density,lebowitz+percus_1961_thermodynamic,hill_1963_thermodynamics,bustamante++_2005_nonequilibrium,seifert_2008_stochastic}. 
In such systems it is not possible to uphold the standard 
interpretation of the thermodynamic quantities
and of their relations. In particular, the  
measurements are made on molecular scales 
and produce values that are dominated 
by fluctuations. Nonetheless, reproducible results are 
obtained by performing averages over many independent 
measurements of small systems at equilibrium 
(e.g., in contact with a heat bath).
Therefore, a proper thermodynamic description is recovered 
if the ensemble method,
which was originally devised as an operational construct to 
obtain results for single macroscopic systems, is given a 
literal interpretation: Thermodynamic quantities are identified
with ensemble averages, which are experimentally realized 
as averages over independent measurements 
under equilibrium conditions. 
Note that the condition of equilibrium requires that 
the small system be in contact with some other system
with a large number of degrees of freedom
(e.g. a heat reservoir for the canonical ensemble),
so that correct results are obtained from the assumption
that the initial state in a particular realization of the process
can be treated as an independent sample from the appropriate 
ensemble. Thus, conditions (i)-(iv) are required 
for the whole and are essential for a proper thermodynamic description
of the small system as well. 

In this work we investigate the transition between 
the regime in which the thermodynamic description is 
valid for a single system and the regime in which 
thermodynamic quantities need to be understood as averages over 
realizations of a given process. To this end we analyze 
the behavior of the empirical estimates 
of equilibrium free energy differences from nonequilibrium work 
measurements by means of the Jarzynski equality.
In contrast to standard thermodynamic relations,
the ensemble average that appears in the Jarzynski equality is 
dominated by rare extreme fluctuations \cite{jarzynski_2006_rare}.
Consequently, to obtain accurate estimations
of the free energy change, one needs to perform repeated 
work measurements in the nonequilibrium process. 
The number of measurements needed to obtain meaningful 
estimates increases exponentially with the 
size of the system \cite{gore++_2003_bias,jarzynski_2006_rare}. 
Therefore, the average
that appears in the Jarzynski equality cannot be realized
in the macroscopic regime, in which definite values of
thermodynamic quantities can be ascribed to single systems 
rather than to collections of systems.

For limited resources, when the number of measurements 
is fixed, the Jarzynski estimator of free energy differences
becomes biased as the size of the system increases. 
In the appropriate scaling limit, the appearance
of the bias is abrupt and corresponds to a phase transition 
in variants of the random energy model (REM) 
\cite{derrida_1980_random,derrida_1981_random,moukarzel+parga_1991_numerical,bovier++_2002_fluctuations,ogure+kabashima_2009_analyticity_1}. 
The connection between the random energy model
and the Jarzynski estimator of free energy differences was made in \cite{palassini+ritort_2008_jarzynski,palassini+ritort_2008_replica,palassini+ritort_2011_improving}. 
In those articles, expressions of the free energy in 
the low-temperature (small $M$ limit, where $M$ is the 
number of nonequilibrium work measurements) and in 
the high-temperature phases (large $M$ limit) 
of the random energy model, including 
finite $M$ corrections, were derived for 
a parametric family of work distributions. 
In another recent work, the convergence of Monte Carlo estimates 
in terms of the random energy model has been made 
in connection with the 'sign problem' \cite{during+kurchan_2010_statistical}. 

In the current paper we further establish the correspondence between the Jarzynski 
estimator of free energy differences and variants of the random energy model 
for the sudden compression of an ideal gas and 
for adiabatic quasi-static volume changes in a dilute real gas.
Even though this correspondence is explicitly established
for only a few particular physical systems, it
is expected to obtain in more general situations.
The origin of the phase transitions in sums of random exponentials is 
the interplay between classic limit theorems for sums 
(e.g., the central limit theorem, the law of large numbers) 
and extreme value statistics 
\cite{benArous_2005_limit,clusel+bertin_2008_global}.
A contribution of this paper is to highlight the importance 
of the phase transition in the Jarzynski estimator of free energy differences
to signal the change between two distinct regimes wherein the ensemble
and the single-system interpretations of thermodynamic quantities 
are applicable, respectively. 

The article is organized as follows: Section II introduces
the Jarzynski equality and related concepts that are necessary 
for the subsequent study. The use of
this equality in the estimation of equilibrium free energy
differences from nonequilibrium work values is described 
in Section III. 
Section IV analyzes the change in behavior of the Jarzynski 
estimator as a function of system size and of the 
number of measurements performed in three illustrative examples. 
Finally, Section V discusses how this change in behavior 
corresponds to the emergence of classical 
macroscopic thermodynamics and a well-defined arrow of time 
as the size of the system is increased,
when the number of measurements performed is fixed. 
The technical details of the analysis of the 
Jarzynski estimator in terms of variants of the random energy 
model are deferred to the appendices.

\section{The Jarzynski equality} \label{sec:JE}
Consider a system characterized by a Hamiltonian 
$H(\Gamma; \lambda)$,  where $\Gamma$ denotes a point 
in phase space. The parameter 
$\lambda$ can be modified by external manipulation
to perform or extract work from the system. If the
system is coupled to other degrees of freedom, 
the changes in $\lambda$ are assumed to affect only 
the Hamiltonian of the 
system of interest and not the terms
that describe its interaction with the environment 
or the environment itself.
 
Assume that the parameter $\lambda$ is modified in the 
interval $[0,\tau]$ according to 
a specified protocol 
$\left\{ \lambda(t); \, 0 \le t \le \tau \right\}$ 
from $\lambda(0) = \lambda_0$ to $\lambda(\tau) = \lambda_1$.  
It is possible to show that the work associated with 
the transition between an initial configuration of the system 
sampled from an equilibrium distribution 
with Hamiltonian $H_0(\Gamma) \equiv H(\Gamma;\lambda_0)$
at temperature $\beta^{-1}$
and a final configuration corresponding to the Hamiltonian 
$H_1(\Gamma)  \equiv H(\Gamma;\lambda_1)$ 
is a random variable $W$ that satisfies the equality 
\cite{jarzynski_1997_nonequilibrium,jarzynski_2004_nonequilibrium}
\begin{equation} \label{eq:jarzynski}
\left< e^{- \beta W} \right> = e^{- \beta \Delta F}, \quad \Delta F = F_1 - F_0.
\end{equation}
The average is performed over trajectories 
whose initial point in phase space 
is sampled from a canonical distribution 
\begin{equation}
\frac{e^{-\beta H_0(\Gamma)}}{\int_{\Omega} d\Gamma e^{-\beta H_0(\Gamma)}},
\end{equation}
where the integral is over $\Omega$, the region of phase space accessible to
the system. This region is assumed to remain unchanged during
the process.
Note that the state in which the system finds itself 
after the manipulation  has been completed need not be an equilibrium one.
In general, $F_1$ is not the free energy of such a state.
Therefore, the free energy difference that appears on the right-hand
side of (\ref{eq:jarzynski})
\begin{equation}
F_i  = - \beta^{-1} \log \int_{\Omega} d \Gamma e^{-\beta H_i(\Gamma)} , \quad i = 0,1,
\end{equation}
need not coincide with the actual free energy change in the process
described \cite{jarzynski_2004_nonequilibrium,palmieri+ronis_2007_jarzynski}. 
It is the difference between the free-energies
of the system in equilibrium at temperature $\beta^{-1}$ in
two states characterized by Hamiltonians $H_0$ and 
$H_1$, respectively.
If the system is strongly coupled with an environment 
that acts as a heat reservoir, $H(\Gamma;\lambda)$
is the potential of mean force associated with 
the variables of the system of interest 
\cite{kirkwood_1935_statistical,jarzynski_2004_nonequilibrium}.
If the system is isolated or
weakly coupled with its environment during the external
manipulation, $H(\Gamma;\lambda)$
can be identified with the Hamiltonian of 
the isolated system \cite{jarzynski_1997_nonequilibrium}.

\section{Estimation
of free energy differences from nonequilibrium work measurements}

Since it was derived 
the Jarzynski equality has attracted a fair amount of interest
because it allows the computation of equilibrium 
free energy differences from measurements of work in general 
nonequilibrium processes
\begin{equation} \label{eq:freeEnergyJarzynski}
\Delta F = - \beta^{-1}  \log \left< e^{- \beta W} \right>,
\end{equation}
where the angular brackets denote an average over the canonical 
ensemble at temperature $\beta^{-1}$ with respect to the initial 
Hamiltoninan $H_0$. 
However, this equality differs in crucial aspects from standard 
thermodynamic relations. 
Because of the molecular nature of the systems analyzed, 
thermodynamic quantities (energy density, number
density, temperature, pressure, work, etc.) are fluctuating variables.
In macroscopic systems, the fluctuations are small. 
This allows us to identify these random thermodynamic quantities
with their typical values, which are deterministic 
and can be measured in single experiments. 
Repetitions of these experiments yield values that are
indistinguishable, within the error of the macroscopic measurement. 
Therefore, for standard thermodynamic quantities, 
typical and average values are close to each other and can be 
used interchangeably to characterize the macroscopic state of 
the system under study.

In contrast, the average that appears in the Jarzynski equality 
needs to be interpreted as a true ensemble average \cite{jarzynski_1997_nonequilibrium}. 
The work measured in a particular realization of a nonequilibrium process
depends on the initial microscopic configuration of the system 
and is therefore a fluctuating quantity.  
In each of these measurements, the system is prepared in an
initial state sampled from the equilibrium distribution.
One then carries out the intervention that gives rise to 
the nonequilibrium process. The work involved in this process
is then recorded. Finally, to extract equilibrium free energy differences
from these measurements, the average of the exponential of minus 
these work values is computed.
Because of the exponential form of the summed quantities, 
typical and average values are, in the general case, very different.
Unlike for standard thermodynamic relations, plugging in the typical value 
of the work on the left-hand side of (\ref{eq:jarzynski}) does not fulfill
the equality. The reason is that the average is dominated by extreme 
events whose probability of occurrence is very low. 
Therefore, a sufficiently large number of measurements is 
needed so that these rare events are well represented and 
the equality can be empirically realized.
The number of measurements required to provide a 
meaningful estimate of the average increases exponentially 
with the size of the system \cite{gore++_2003_bias,lua+grosberg_2005_practical,jarzynski_2006_rare}. 
Therefore, it is not practicable
in macroscopic systems. Nonetheless, the regime in which the
equality can be experimentally verified is accessible 
for sufficiently small systems. 

To realize the average, repeated experiments under
the same conditions are carried out. The values of 
work obtained in each of these experiments are recorded
and used to estimate the average
\begin{equation}
\left< e^{- \beta W} \right>  \approx 
\left< e^{- \beta W} \right>_M \equiv \frac{1}{M} \sum_{m=1}^M  e^{- \beta W_m}.
\end{equation}
By means of the Jarzynski equality, this Monte Carlo 
average can be used to estimate the change in free energy 
\begin{equation}
\Delta F_M \equiv - \beta^{-1}  \log \left< e^{- \beta W} \right>_M.
\end{equation}
Since the particular realization of work values $\left\{W_m \right\}_{m=1}^M$
is random, the estimate $\Delta F_M$ is also a random variable.
Our goal is to understand the properties of this random
variable as a function of $M$, the number of nonequilibrium
work measurements and $N$, the size of the system.
 
Because of the exponential form of the quantity averaged, the role
of extreme fluctuations is very important. In contrast 
to usual thermodynamic averages, such as $\left<W\right>$, 
the average work, which are dominated by configurations that are typical
of the initial (equilibrium) state of the system, 
the average $\left< e^{- \beta W} \right>$
is actually dominated by rare configurations that are
typical of the system at equilibrium at temperature $\beta^{-1}$ 
with respect to the {\it final} Hamiltonian \cite{jarzynski_2006_rare}.

Of particular interest is the block \cite{zuckerman+woolf_2002_theory,zuckerman+woolf_2004_systematic} 
or quenched  \cite{mezard+montanari_2009_information} average
\begin{equation} \label{eq:quenchedF}
\mathbb{E}\left[\Delta F_M \right] \equiv 
- \beta^{-1} \mathbb{E}\left[ \log \left< e^{- \beta W} \right>_M \right],
\end{equation}
where the expectation $\mathbb{E}\left[\cdot \right]$ 
is with respect to independent realizations of $M$  measurements,
each of which corresponds to an independent sample from an initial 
equilibrium canonical distribution at temperature $\beta^{-1}$. 
In \cite{zuckerman+woolf_2002_theory,zuckerman+woolf_2004_systematic}
$\mathbb{E}\left[\Delta F_{M} \right]$ 
is referred to as the {\it finite-data average free energy}. 
This quantity is an estimator of $\Delta F$, albeit 
a biased one, in general. 
The bias is the difference between the expected value of this estimator 
and the actual value of the free energy
\begin{equation}
B_{M}  \equiv  \mathbb{E}\left[\Delta F_{M} \right]  -  \Delta F. 
\end{equation}
Using the law of large numbers it is possible to show 
that in the limit $M \rightarrow \infty$ the 
block average converges to the free energy change 
\begin{equation} \label{eq:DeltaF_inf}
\lim_{M \rightarrow \infty} \mathbb{E}\left[ \Delta F_{M} \right] 
=  \Delta F. 
\end{equation}
Therefore, the estimator $\mathbb{E}\left[\Delta F_{M}\right]$ 
is asymptotically unbiased
\begin{equation}
\lim_{M \rightarrow \infty}  B_{M}  =  0.
\end{equation}

For a single experiment $M=1$ the estimator equals the 
average work performed on the system
\begin{equation}
\mathbb{E}\left[\Delta F_{1} \right] = \left< W \right>.
\end{equation}
Following \cite{gore++_2003_bias}, we define the `dissipated work' 
in a given realization of the experiment as   
\begin{equation}
W_{dis} = W - \Delta F.
\end{equation}
Using Jensen's inequality it is possible to show that
$\mathbb{E}\left[\Delta F_{1}\right]$ is a positively biased estimator 
of $\Delta F$
\begin{equation}
\left< W \right> = \mathbb{E}\left[\Delta F_{1} \right] \ge \Delta F.
\end{equation}
In fact, the convergence of $\mathbb{E}\left[\Delta F_{M}\right]$
to the asymptotic limit $\Delta F$ is monotonic 
\cite{zuckerman+woolf_2004_systematic}
\begin{equation}
\left< W \right> = \mathbb{E}\left[\Delta F_{1}\right] \ge \mathbb{E}\left[\Delta F_{M}\right] \ge  \mathbb{E}\left[\Delta F_{M+1}\right] \ge
\mathbb{E}\left[\Delta F_{\infty}\right] = \Delta F, \quad 1 < M < \infty.   
\end{equation}
In terms of the bias
\begin{equation}
\left< W_{dis} \right> =  \left< W \right> - \Delta F 
= B_{1} \ge  B_{M} \ge B_{M+1} \ge B_{\infty}  = 0, 
\quad 1 < M < \infty.  
\end{equation}
Therefore, the maximum bias corresponds to $M=1$ and coincides with
the average dissipated work  
\begin{equation}
B_{\max} = B_{1}  = \left< W_{dis} \right>.
\end{equation}

The main contribution of this research is
to explicitly show in several 
paradigmatic cases that the finite sample estimate of free energy 
differences from the Jarzynski equality
for a particular system exhibits a change of 
behavior as $M$, the number of repetitions of the experiment, 
increases. Alternatively, for a fixed number of measurements, 
the regime change occurs as a function of $N$, the system size.
For small systems
the Jarzynski estimator is unbiased. In these systems
the nonequilibrium work measurements are dominated 
by fluctuations.
Configurations that are typical of the reversed
process are well sampled, which 
means that the arrow of time is poorly defined.
As the system size increases, the probability
of sampling these configurations becomes exponentially
small, so that they are not observed in practice.
The Jarzynski estimator of the free
energy change becomes biased and asymptotically
approaches the value of the average work. 
The suppression of these fluctuations also leads to the
emergence of a well-defined arrow of time in the system. 
In the limit $M \rightarrow \infty$, $N \rightarrow \infty$
with $\log M / N \rightarrow \text{constant}$
the regime change is akin to a phase transition 
that arises in simplified models of
spin-glasses, such as the random energy model 
in both its continuous 
\cite{derrida_1980_random,derrida_1981_random} 
and discrete \cite{moukarzel+parga_1991_numerical,moukarzel+parga_1992_numerical,ogure+kabashima_2009_analyticity_1} versions.

\section{Phase transition in the Jarzynski estimator 
of free energy differences}

We now proceed to analyze the behavior of the Jarzynski estimator
in three important cases. The first one corresponds
to processes in which the nonequilibrium work distribution 
is Gaussian \cite{gore++_2003_bias}. This is a particular
case of the class of work distributions analyzed in 
\cite{palassini+ritort_2011_improving} with $\delta = 2$.
The second case is a compression experiment for an ideal gas 
\cite{lua+grosberg_2005_practical,presse+silbey_2006_ordering,palmieri+ronis_2007_jarzynski}. 
Finally, we consider adiabatic and quasi-static volume changes
for a dilute classical gas of interacting particles \cite{crooks+jarzynski_2007_work}.

The change of regime is best analyzed in terms of the normalized bias
\begin{equation}
\tilde{B}_{M} \equiv
 \frac{B_{M}}{B_{\max}} 
= \frac{\mathbb{E}\left[\Delta F_{M}\right]  -  \Delta F}{\left< W \right>  -  \Delta F}, \quad  
0 \le \tilde{B}_{M} \le 1,
\end{equation}
where the maximum bias is the difference between the average work 
in the actual nonequilibrium process 
(which is not necessarily isothermal)
and the free energy difference in the corresponding
isothermal process
\begin{equation}
B_{\max} = \left< W \right> - \Delta F. 
\end{equation}
As a function of $M$, the normalized bias is a 
monotonically decreasing quantity of $M$, which is 
bounded between $0$ and $1$
\begin{equation}
1 = \tilde{B}_{1} \ge \tilde{B}_{M}\ge \tilde{B}_{M+1} \ge \tilde{B}_{\infty}  = 
0, \quad 1 < M < \infty.  
\end{equation}
To simplify the derivations we assume $\beta = 1$. 
It is straightforward to reintroduce 
this parameter in the final expressions 
by noting that setting $\beta = 1$ is equivalent
to measuring energies in units of $\beta^{-1} = k_B T$, where
$k_B$ is the Boltzmann constant and $T$ is the initial equilibrium 
temperature.

\subsection{Gaussian work distribution \label{sec:REM}} 

In this section, we illustrate the connection 
between the Jarzynski free energy difference estimator and the random
energy model for a Gaussian work distribution. The results presented
in this section were first derived in \cite{palassini+ritort_2011_improving}.
That reference gives explicit expressions for the bias 
of the Jarzynski estimator in different regimes
for a general class of work distributions, which includes the Gaussian
as a particular case. It also considers finite-size corrections,
which are ignored in our analysis.

Assume that in the sample estimate
\begin{equation}
\left< e^{- W} \right>_M = \frac{1}{M} \sum_{m=1}^M  e^{- W_m}.
\end{equation}
the  work values  $\left\{ W_m \right\}_{m=1}^{M}$ follow a normal 
distribution whose mean is  $< W >$, and whose variance is  $\sigma^2$ 
\cite{gore++_2003_bias}. In this case,
\begin{equation}
\Delta F =  - \log \left< e^{- W} \right>  = \left< W \right> - \frac{1}{2} \sigma^2.
\end{equation}
Therefore, the  maximum bias is
\begin{equation}
B_{\max} = \left< W \right> - \Delta F =  \frac{1}{2} \sigma^2.
\end{equation}
This is an extensive quantity and scales with the
size of the system. In the limit $\sigma \rightarrow \infty$,
$M \rightarrow \infty$ and $ \log M / \sigma^2 $ finite,
the estimate of the free energy 
has an abrupt change of behavior (see appendix \ref{app:REM})
\begin{equation} \label{eq:DeltaF_M}
\mathbb{E}\left[\Delta F_{M}\right] 
= \left\{ 
\begin{array}{ll}
\left< W \right> - \frac{1}{2} \sigma^2, 
 &  M \ge \exp \left\{\frac{1}{2} \sigma^2 \right\} \\
\left< W \right> -\sqrt{2} \sigma \sqrt{\log{M}}  + \log M, &
 M  < \exp \left\{\frac{1}{2}\sigma^2 \right\}.
\end{array}
\right. 
\end{equation}
These expressions correspond to those derived 
in \cite{palassini+ritort_2011_improving} 
(Eq. (4) and the paragraph before this equation
in that reference) 
for $\delta = 2, $ $\Omega^2 = 2 \sigma^2,$ $N = M$,
and $D_c = \sigma^2/2$. 

For a fixed value of $\sigma$ the normalized bias is
\begin{equation}  \label{eq:bias_M_M}
\tilde{B}_M = \frac{B_M}{B_{\max}} =
\left\{ 
\begin{array}{ll}
 0,
 &  M \ge  M_c \\
\left(1 - \sqrt{\log M / \log M_c }  \right)^2, &
 M  < M_c
\end{array}
\right. 
\end{equation}
with $M_c = \exp \left\{\frac{1}{2} \sigma^2 \right\}$.

\begin{figure}
\includegraphics[scale=0.7]{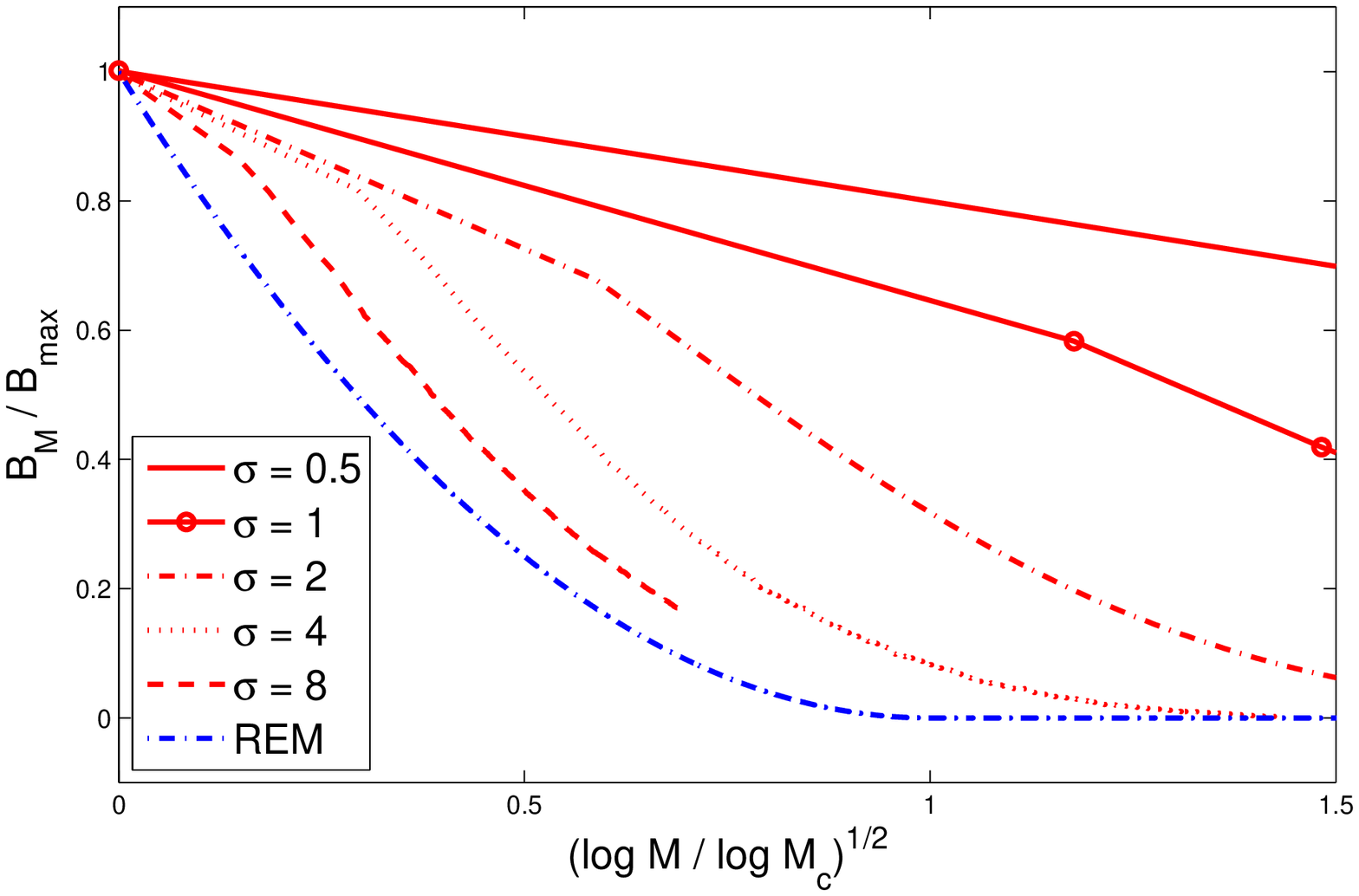}
\caption{
Approach of the normalized bias to the curve 
(\ref{eq:bias_M_M}) as $M\rightarrow \infty$.
The random energy model corresponds to the 
dash-dotted line.
}\label{fig:bias_M_M}
\end{figure}

Figure \ref{fig:bias_M_M} displays the
dependence of the normalized bias 
(\ref{eq:bias_M_M}) as a function of
$\sqrt{\log M / \log M_c}$ with a fixed
$\sigma$, for different values of $\sigma$. 
The curve corresponding to the asymptotic limit
$\sigma\rightarrow \infty$ is plotted as a dash-dotted line.
The remaining curves are averages over Monte Carlo
simulations. 
For small numbers of measurements ($M < M_c$), 
the Jarzynski estimator is biased. In this regime, 
the free energy exhibits strong (of order 1) non-Gaussian 
fluctuations around its average
(Theorem 1.6 from \cite{bovier++_2002_fluctuations}). 
These fluctuations 
are driven by the Poisson process of the extremes
of the random nonequilibrium work measurements.
The range $M > M_c$ corresponds to a regime in 
which the estimate of the free energy change
for sufficiently large $\sigma$ is unbiased.
The bias persists beyond this limit 
for small systems: when $\sigma$ is small,
the transition between regimes is more
gradual. This is the region in which one expects 
to observe convergence to the Jarzynski limit in 
experiments.  
There is a second phase transition
in the system: in the range
$  e^{\sigma^2/2} < M < e^{\sigma^2} $
the fluctuations can be expressed in terms
of the Poisson process of extremes
of the nonequilibrium work measurements. Beyond
the threshold $ M_c^{'} = e^{\sigma^2} $, 
the central limit theorem holds and the fluctuations 
around the block average are approximately Gaussian 
(Theorem 1.5 from \cite{bovier++_2002_fluctuations}).
The graph presented corresponds to 
Fig. 2(a) in \cite{palassini+ritort_2011_improving},
with $\sigma^2 = \Omega^2 /2 $. The main difference
is the square root in the abscissae, which does not
modify the point at which the phase transition occurs in the 
REM limit or the qualitative picture.

Table \ref{tab:table1} displays the 
number of measurements needed to obtain an unbiased
estimate of the free energy using the Jarzynski estimator ($M_c$) 
and to reach a regime in which the fluctuations around 
this estimate are Gaussian ($M_c^{'}$) for the several
values of $B_{max}$ in the range
explored in the experiments
described in \cite{palassini+ritort_2011_improving} 
(from $k_B T$ to $20 k_B T$).
 
\begin{table}[b]
\caption{\label{tab:table1}%
Number of measurements needed to obtain an unbiased estimate
of the free energy ($M_c$)
and to reach a regime in which the fluctuations around
this estimate are Gaussian ($M_c^{'}$)
as a function of $B_{max}$, when the nonequilibrium work distribution is
Gaussian.}
\begin{ruledtabular}
\begin{tabular}{lccccc}
\textrm{$B_{max} $}&
\textrm{$ k_B T$}&
\textrm{$2 k_B T$}&
\textrm{$5 k_B T$}&
\textrm{$10 k_B T$}&
\textrm{$20 k_B T$}\\
\colrule
$M_c $     & 3 &  8 &  149   &    22,027     &  485,165,196 \\
$M_c^{'}$  & 8 & 55 & 22,027  &   485,165,196 & $2.35 \cdot 10^{17}$
\end{tabular}
\end{ruledtabular}
\end{table}

The regime change can also be observed
when the number of measurements is fixed
and the size of the system, measured
in terms of $ \sigma^2$, increases.
For a fixed $M$, the normalized bias is
\begin{equation}  \label{eq:bias_M_sigma}
\tilde{B}_M = \frac{B_M}{B_{\max}} =
\left\{ 
\begin{array}{ll}
 0,
 &  \sigma \le  \sigma_c \\
\left(1 - \sigma_c / \sigma \right)^2, & \sigma >  \sigma_c
\end{array}
\right.
, 
\end{equation}
where $ \sigma_c = \sqrt{2 \log M}$.

\begin{figure}
\includegraphics[scale = 0.7]{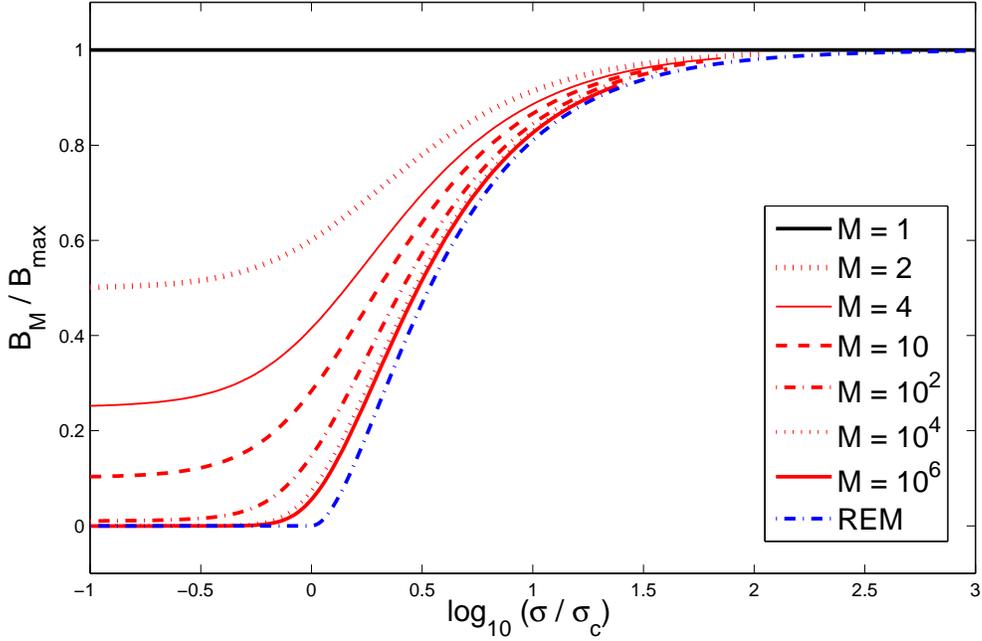}
\caption{
Approach of the normalized bias to the curve 
(\ref{eq:bias_M_sigma}) as $M\rightarrow \infty$.
The random energy model corresponds to the dash-dotted curve.
}\label{fig:bias_M_sigma}
\end{figure}

Figure \ref{fig:bias_M_sigma} displays the
curves that trace the dependence of the normalized bias 
(\ref{eq:bias_M_sigma}) as a function of
$\log_{10}(\sigma / \sigma_c)$ with
$M$ fixed, for different values of $M$. 
The curve corresponding to the random energy model
($M\rightarrow \infty$) is displayed as a dash-dotted
line.
In small systems $\sigma < \sigma_c$  
the estimate of the free energy difference
is unbiased.
For sufficiently small systems ($\sigma \rightarrow 0$), 
the bias scales as $B_M \sim B_{\max}/M$. Therefore, 
the bias is reduced by increasing $M$ linearly. 
This is the region in which one expects 
convergence of the sample average
to the Jarzynski equality limit.
Beyond that threshold ($\sigma > \sigma_c$), 
the empirical estimate is
biased. As the system size increases the bias approaches
its maximum value, which corresponds to measuring the
typical value in most of the nonequilibrium work measurements,
in agreement with the expected behavior of classical 
macroscopic systems. In this regime, linear increases in 
$M$ do not lead to significant changes in the bias observed.

\subsection{Ideal gas compression experiment \label{sec:cDREM}} 
Consider an isolated system consisting of
$N$ non-interacting particles (ideal gas).
The system is confined in the interval $[-L,L]$ in the X direction. 
The macroscopic state of the system is defined by the
temperature and the value of an externally
applied potential.
The microscopic state of the system is 
characterized by $n$, the number of particles
in region II ($0 < x \le L$). 
Correspondingly, the number of number of 
particles in region I ($-L \le x \le 0$) is $N-n$. 

Initially, the external potential is zero and 
the system is assumed to be in a homogeneous
equilibrium state at temperature $\beta^{-1} = 1$. 
The probability of having $n$ particles in region 
II in this state is
\begin{eqnarray}
p(n;0) & = & \frac{1}{2^N} \binom{N}{n}, 
\quad  n = 0,1,\ldots,N.
\end{eqnarray} 
This is a binomial distribution with the parameters
$(N,1/2)$. 
It is peaked around the mean $n^*(0) = N/2 $. Its 
standard deviation is  $\sqrt{N}/2$.
The corresponding free energy is
\begin{equation}
F_N(0) = - \log \left(\frac{1}{2^N} \sum_{n=0}^N  
\binom{N}{n} \right) = 0.
\end{equation}

Consider a compression experiment during which 
the system undergoes a sudden transition from the initial 
homogeneous equilibrium state to a state in which particles 
are more likely to be in region I because of the presence
of a positive external potential in region II
\begin{equation}
V(x; \epsilon)  = \left\{ \begin{array}{lrlll}
                              0,        &    -L & \le  & x \le 0  \quad \hbox{[region I]}\\
                             \epsilon,  &     0 & < &x \le L \quad \hbox{[region II]}
                          \end{array}
                   \right..
\end{equation}
The equilibrium distribution 
at temperature $\beta^{-1} = 1$, when the system is subject to
this external potential is 
a binomial distribution with parameters 
$\left(N, e^{-\epsilon}/(1+e^{-\epsilon}) \right)$
\begin{eqnarray}
p(n;\epsilon) & = & 
 \frac{1}{\left(1+e^{-\epsilon}\right)^{N}} 
\binom{N}{n}
e^{-n\epsilon}, \quad n = 0,1,\ldots,N.
\end{eqnarray}
This distribution is peaked around its mean
$n^*(\epsilon) = N/(1+e^{\epsilon})$. 
Its standard deviation is 
$\sqrt{N} e^{-\epsilon/2} / \left(1+e^{-\epsilon}\right)$.
The corresponding free energy is
\begin{equation}
F_N(\epsilon) = - \log \left[\frac{1}{2^N} \sum_{n=0}^N  
\binom{N}{n}
 e^{-n\epsilon} \right]  = 
 N \log \frac{2}{1+ e^{-\epsilon}}.
\end{equation}

The change in free energy between the initial state, in 
which the system is in equilibrium at temperature 
$\beta^{-1} = 1$, in the absence of a external potential,
and an equilibrium state at the same temperature
with potential $V(x; \epsilon)$ is
\begin{equation} \label{eq:deltaFidealGasCompression}
\Delta F = F_N(\epsilon) - F_N(0) = N \log \frac{2}{1+ e^{-\epsilon}}.
\end{equation}
As noted by several authors, this is not the change of free energy between the equilibrium states corresponding to the initial and final values of the work parameter in the experiment that is being performed
\cite{jarzynski_2004_nonequilibrium,palmieri+ronis_2007_jarzynski}.
In particular, the configuration of the particles in the system at $t = 0^+$  
is the same as in the initial state because there has not been 
any time to evolve. In most cases, this configuration
is very atypical of the equilibrium state for 
the new constraints. In fact, the system does not have 
a well-defined temperature in this state. 
Nonetheless, using the equality 
\begin{equation}
\Delta F = - \log \left<e^{-W}\right>,
\end{equation}
it is possible to compute
the change in free energy in an isothermal process
(\ref{eq:deltaFidealGasCompression}) in
terms of the external work performed on the system
during the compression \cite{presse+silbey_2006_ordering}.

In practice, one needs to carry out a series of independent 
realizations of  the experiment. In the $m$th realization, 
the configuration of the system is sampled from 
the equilibrium distribution $ n_m \sim p(n;0)$.
The external work needed to perform the transition
in this particular realization is $W_m = n_m \epsilon$.
The average work over $M$ realizations of the experiment is 
\begin{equation}
\left< W \right>_M = \frac{1}{M} \sum_{m=1}^M  W_m = 
\frac{1}{M} \sum_{m = 1}^M  n_m \epsilon.
\end{equation}
Since the $\left\{n_m \right\}_{m=1}^M$ are independent
identically distributed random variables (iidrvs),
$\left< W \right>_M$ is also a random variable whose average is 
the thermodynamic work
\begin{equation}
\left< W \right> = \mathbb{E}\left[ \left< W \right>_M \right] = 
\frac{1}{M} \sum_{m = 1}^M  \mathbb{E}\left[ n_m \right] \epsilon
=  \mathbb{E}\left[ n \right] \epsilon
= \frac{1}{2} N \epsilon.
\end{equation}
The average and the typical value of the work performed
coincide.

Consider now the Jarzynski estimator of the free-energy difference 
\begin{equation}
\Delta F_M = - \log \left<e^{-W} \right>_M = 
- \log \left(\frac{1}{M} \sum_{m=1}^M  e^{-W_m} \right) =
- \log \left(
\frac{1}{M} \sum_{m = 1}^M  e^{-n_m \epsilon} \right).
\end{equation}
This is also a random variable whose average is
\begin{equation}
\mathbb{E}\left[\Delta F_{M}\right] = - \mathbb{E} \left[\log \left< e^{-W} \right>_M \right] = 
- \mathbb{E} \left[\log \left(\frac{1}{M} \sum_{m=1}^M e^{- W_m} \right) \right] = - \mathbb{E} \left[\log \left(
\frac{1}{M} \sum_{m = 1}^M  e^{-n_m \epsilon} \right)\right].
\end{equation}
For $M = 1$ this average coincides with the average work
\begin{equation}
\mathbb{E}\left[\Delta F_{1} \right] = \left< W \right>  = \frac{1}{2} N \epsilon.
\end{equation}
In the limit $M \rightarrow \infty$, it converges to 
the free energy change 
\begin{equation}
\lim_{M \rightarrow \infty} \mathbb{E}\left[\Delta F_{M}\right] = \Delta F = 
N \log \frac{2}{1+ e^{-\epsilon}}.
\end{equation}

There is no closed-form expression for this average 
for other values of $M$. 
However, taking advantage of 
the correspondence between the ideal gas compression experiment 
and the discrete random energy model (see appendix \ref{app:DREM}),  
it is possible to show that there is a continuous 
but abrupt change in the block average
\begin{equation}
\mathbb{E}\left[\Delta F_{M}\right] = \left\{ 
\begin{array}{ll} 
N \log \frac{2}{1+e^{-\epsilon}}, & \epsilon  \le \epsilon_c(\gamma) \\
N \left[\gamma \log 2 +  \frac{\epsilon}{2} \left( 1 - \tanh \frac{\epsilon_c}{2}  \right) \right]
= N \left[ \gamma \log 2 + \frac{\epsilon}{1+e^{\epsilon_c}} \right],  
& \epsilon   > \epsilon_c(\gamma) 
\end{array}
\right.
\end{equation}
in the limit $M \rightarrow \infty$, 
$N \rightarrow \infty$, 
with 
\begin{eqnarray}
\gamma & = & \frac{\log M} {N \log 2}\rightarrow \hbox{constant}, \\
\epsilon_c(\gamma) & = & \left\{
\begin{array}{ll} 
 \infty, \quad & \gamma  \ge 1 \\
 \log \left[1- h_2^{-1}(1-\gamma) \right] - \log \left[ h_2^{-1}(1-\gamma) \right], \quad & \gamma  < 1  
\end{array}
\right. .
\end{eqnarray}
The function $h_2^{-1}(y) \in [0, 1/2]$ is the inverse of the binary entropy
\begin{equation}
h_2(x) =  -x \log_2 x -(1-x) \log_2(1-x).
\end{equation}
In this limit the bias of $\mathbb{E}\left[\Delta F_{M}\right]$ 
as an estimator of $\Delta F$ is
\begin{equation}
B_{M} = \mathbb{E}\left[\Delta F_{M}\right] - \Delta F = \left\{ 
\begin{array}{ll}
0, &  \epsilon \le \epsilon_c(\gamma) \\
N \left[\gamma \log 2 +  \frac{\epsilon}{2} \left( 1 - \tanh \frac{\epsilon_c}{2}  \right) - \log \frac{2}{1+e^{-\epsilon}} \right],   & \epsilon > \epsilon_c(\gamma) 
\end{array}
\right. .
\end{equation}

For a fixed value of $\epsilon$ the bias is maximum for  
$M=1$ ($\gamma = 0$, $ \epsilon_c = 0$)
\begin{equation}
B_{\max}(\epsilon) = B_{1} = N \left[ \frac{\epsilon}{2}  - \log \frac{2}{1+e^{-\epsilon}} \right]  = N \log \cosh \frac{\epsilon}{2}.
\end{equation}
The normalized bias is
\begin{equation}
\tilde{B}(\gamma,\epsilon) = \frac{B_{M}}{B_{\max}(\epsilon)} = 
\left\{ 
\begin{array}{ll}
0, &  \gamma \ge \gamma_c(\epsilon) \\
1 - \frac{\frac{\epsilon}{2} \tanh\frac{\epsilon_c}{2} - \gamma \log 2}{
\log \cosh \frac{\epsilon}{2}} =
1 -  \frac{\epsilon - \epsilon_c}{2}\frac{\tanh\frac{\epsilon_c}{2}}{
\log \cosh \frac{\epsilon}{2}}
- \frac{\log \cosh \frac{\epsilon_c}{2}}{
\log \cosh \frac{\epsilon}{2}}, 
 & \gamma < \gamma_c(\epsilon) 
\end{array}
\right.
,
\end{equation}
where 
\begin{equation}
\gamma_c(\epsilon) = \epsilon_c^{-1}(\epsilon) = 1 - h_2 \left(\frac{1}{1+e^{\epsilon}} \right),
\end{equation}
with $\gamma_c(0) = 0$ and $\lim_{\epsilon \rightarrow \infty} \gamma_c(\epsilon) = 1$.

\begin{figure} 
\includegraphics[scale=0.41]{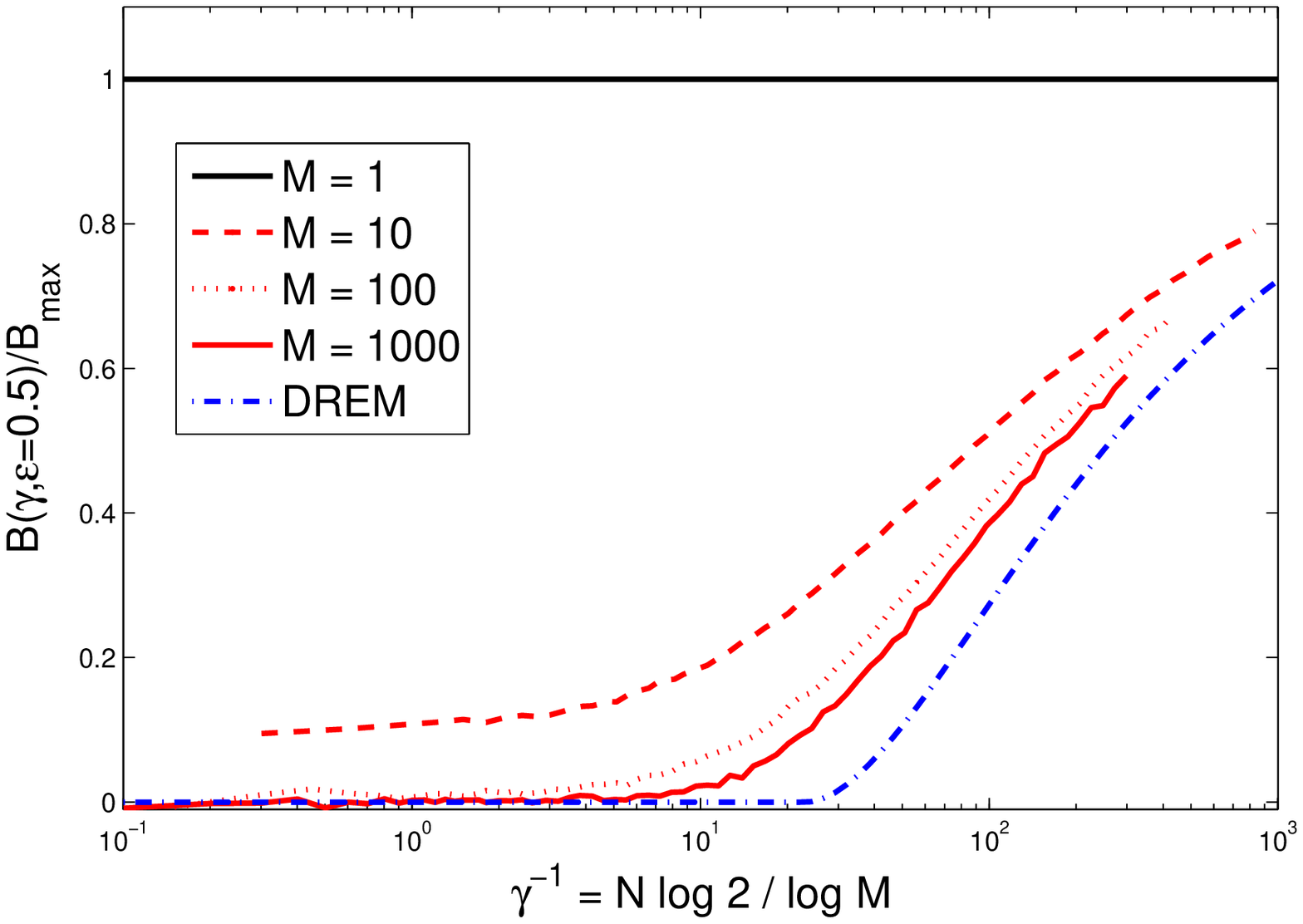} 
\includegraphics[scale=0.41]{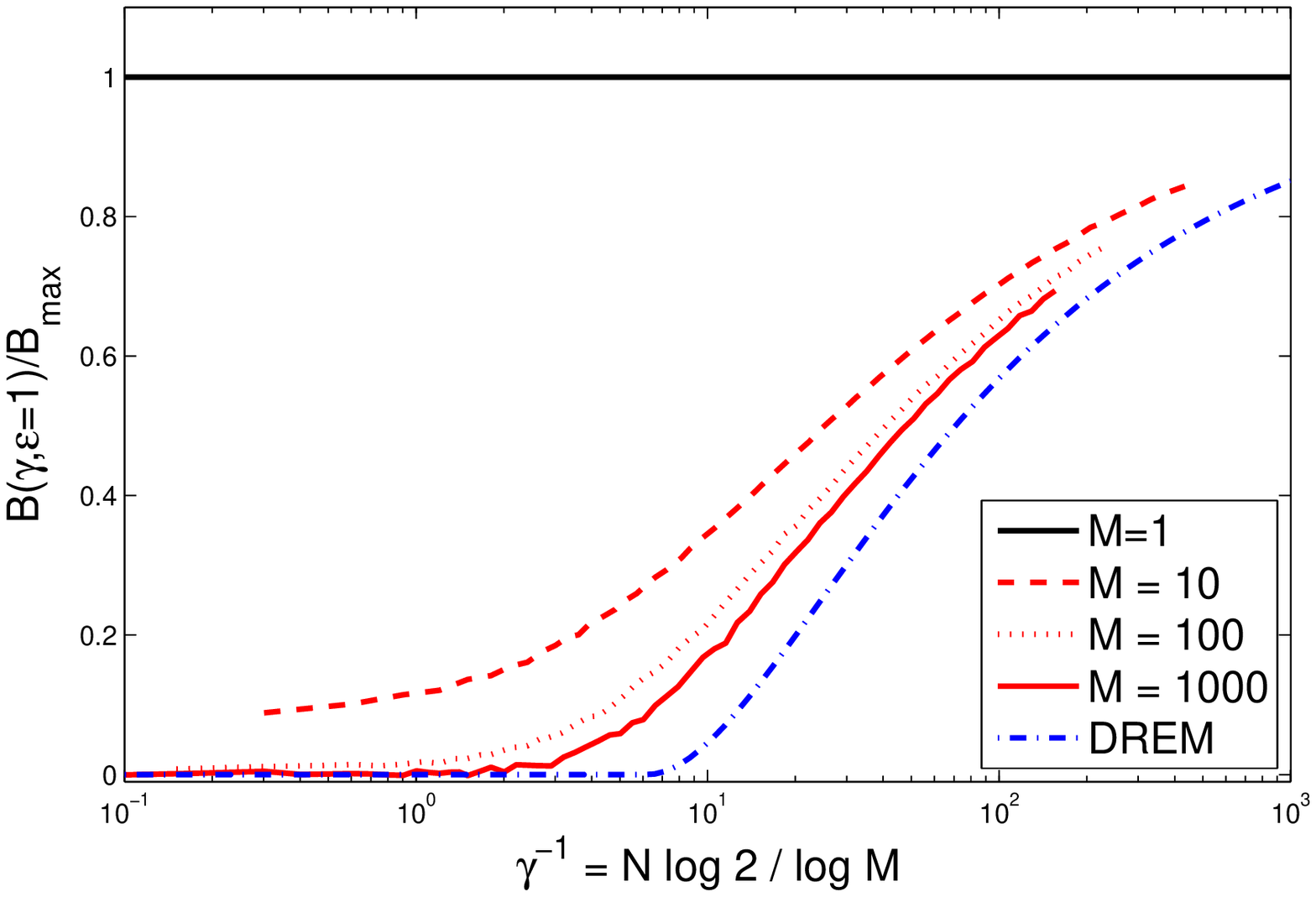}   
\includegraphics[scale=0.41]{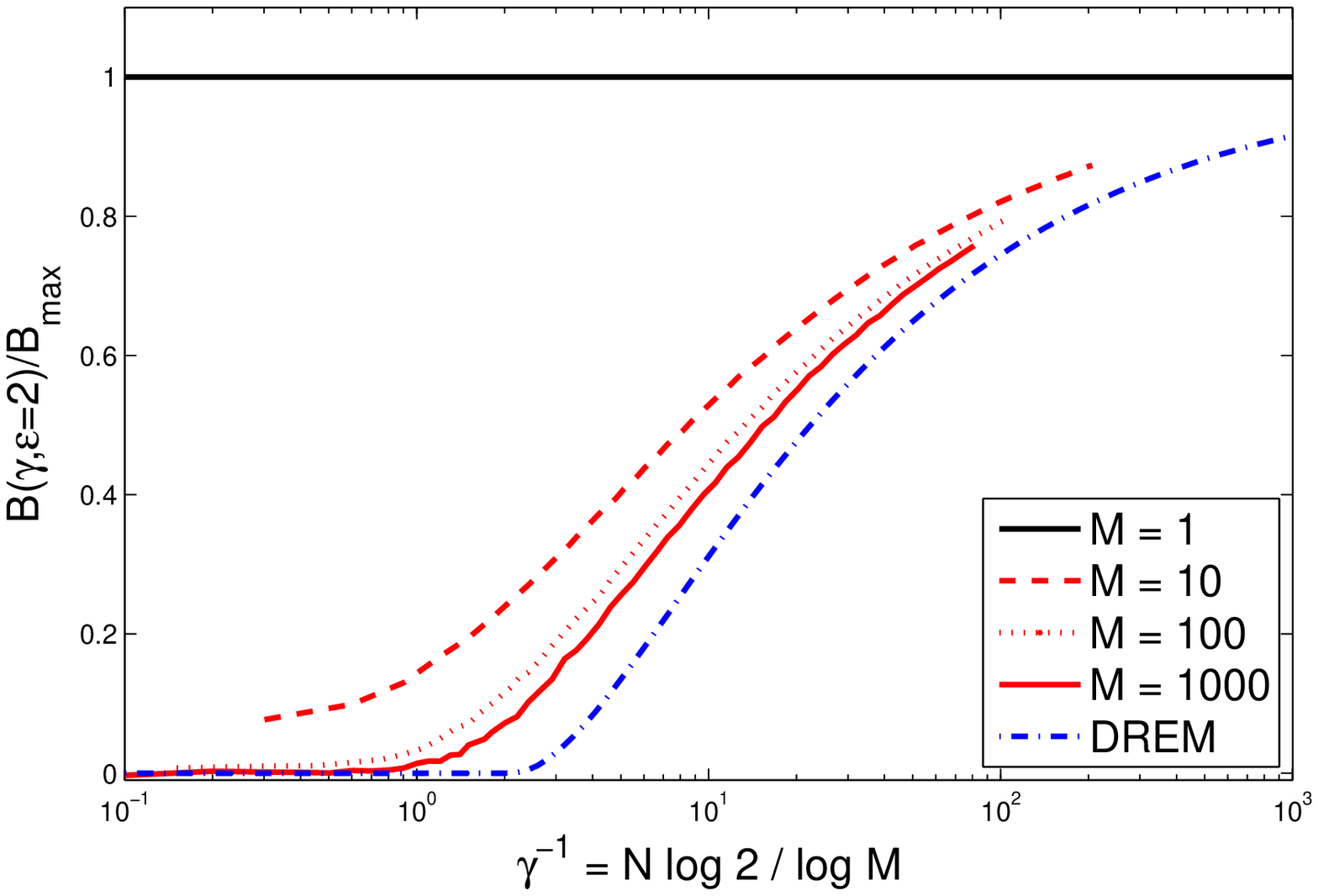}   
\includegraphics[scale=0.41]{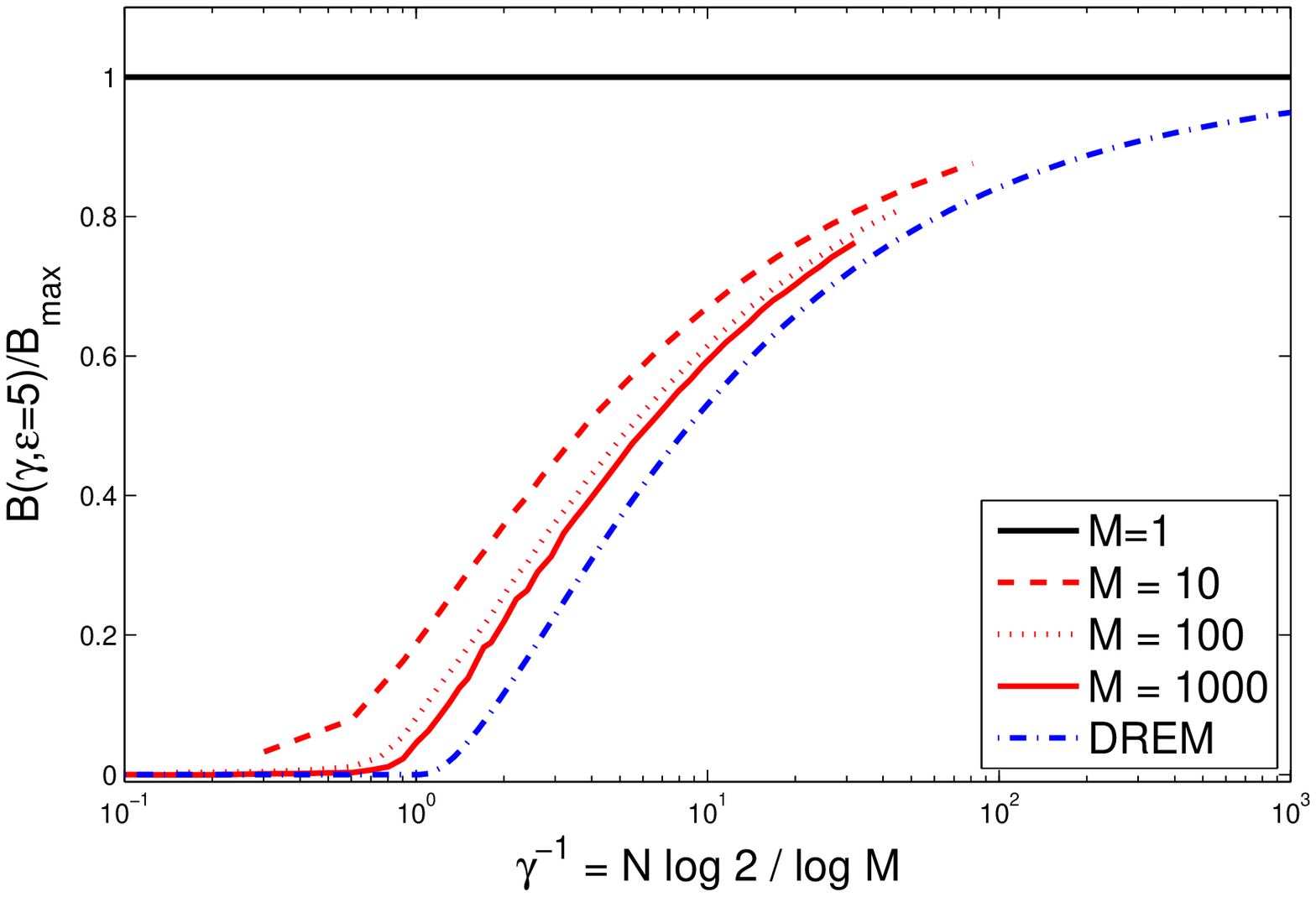} 
\caption{Dependence of the normalized bias on $N$ for 
$\epsilon = 0.5$ (top left), $\epsilon = 1$ (top right), 
$\epsilon = 2$ (bottom left) and $\epsilon = 5$ (bottom right).
The asymptotic value for the discrete random energy model 
corresponds to the dash-dotted curve in the plots.
}\label{fig:bias_DREM_N_epsilon}
\end{figure}

The results of computer simulations of the ideal gas
compression experiment at temperature $\beta^{-1}  = 1$
are depicted in figure \ref{fig:bias_DREM_N_epsilon}.
The graphs display, for different values of $\epsilon$, the dependence
of the normalized bias as a function of system size, measured
in terms of $\gamma^{-1}$ with $M$ fixed, for different values of $M$.
The behavior observed is similar to the Gaussian case: the regime
in which the Jarzynski equality can be realized (i.e. the Jarzynski 
free energy estimator is unbiased) corresponds to gases composed of
a small number of particles. If the system size is increased, 
assuming that the number of measurements is kept fixed, a 
gradual change takes place to a regime in which the Jarzynski estimator becomes
biased. Asymptotically, for large $N$, the estimator becomes
maximally biased, which means that only typical work values 
are observed. Linear increases in $M$ do not significantly reduce
this bias. Therefore, in this regime, it is not possible to 
empirically realize the Jarzynski equality and
measurements yield standard macroscopic thermodynamic values.
The change in regime becomes more abrupt as $\epsilon$, $M$ and
$N$ increase and is asymptotically well described by 
the phase transition that takes place in the discrete random
energy model. 

Table \ref{tab:table2} displays the 
number of particles ($N_c$) at which the transition from
a regime in which the Jarzynski estimator is 
unbiased ($N < N_c$) to a regime in which the Jarzynski estimator is 
biased ($N > N_c$), for several values of $M$
and $\epsilon$. Analyzing the results presented in 
this table one can see that
the critical system size is rather small and increases 
rather slowly (logarithmically) with $M$, the number of 
work measurements performed.

\begin{table}[b]
\caption{\label{tab:table2}%
Number of particles ($N_c$) at which the transition between
the two regimes of the Jarzynski estimator occurs 
for different values of $M$, the number of 
work measurements performed,
and $\epsilon$, the value of the external potential applied
in the compression of an ideal gas. 
}
\begin{ruledtabular}
\begin{tabular}{lcccc}
\textrm{}&
\textrm{$\epsilon = 0.5$}&
\textrm{$\epsilon = 1$}&
\textrm{$\epsilon = 2$}&
\textrm{$\epsilon = 5$}\\
\colrule
M = 10     &  76 & 21 &  8 & 4\\
M = 100    & 152 & 42 & 15 & 8\\
M = 1000   & 228 & 63 & 22 & 11\\
M = 10,000 & 304 & 84 & 29 & 15\\
\end{tabular}
\end{ruledtabular}
\end{table}

\subsection{Adiabatic quasi-static volume change in a dilute gas.} 
\label{subsec:adiabatic}

Consider a dilute gas of $N$ interacting particles in $d$ dimensions. 
Assume that this gas undergoes an adiabatic quasi-static volume change 
from $V_0$ to $V_1$. The work distribution in this process is
\begin{equation}
p(W) =  \frac{1}{\left|\alpha\right|\Gamma(K)}
\left(\frac{W}{\alpha}\right)^{K-1}  e^{- W / \alpha} \theta(\alpha W),
\end{equation}
where $K = N d/2$ and $\alpha = (V_0/V_1)^{2/d}-1$ \cite{crooks+jarzynski_2007_work}. 
The Heaviside step function $\theta(\alpha W)$ guarantees
that work is positive for compression ($\alpha > 0$) and negative 
for expansion ($\alpha < 0$).
The average work is $\left< W \right> = (Nd/2) \alpha$ . The typical value can be 
identified with the mode of the distribution $W_{typ} =(Nd/2-1) \alpha$
for $Nd/2 >1$. Note that for large systems ($N\rightarrow \infty$) typical
and average work values are very similar.

Assuming a fixed number of measurements $M$ there is 
an abrupt change of behavior of the Jarzynski estimator as
$N$, the size of the system, increases, for
sufficiently large $M$ and $N$ (see Appendix \ref{app:Gamma_REM})
\begin{eqnarray}
\mathbb{E}\left[\Delta F_{M}\right] & = & -\mathbb{E} \left[\log \left< e^{-W}\right>_M \right] =
\left< W \right> + \log M - \mathbb{E} \left[\log \sum_{m=1}^M 
e^{- \alpha E_m} \right]  \nonumber \\
& = & 
\left\{
\begin{array}{ll}
 N \frac{d}{2} log(1+\alpha),                 & N \le N_c \\
 N \frac{d}{2} \alpha (1+x_l(N)) + \log M,       & N  > N_c  
\end{array}
\right.
,
\end{eqnarray}
where the system size at which the transition takes place is
\begin{equation}
N_c = \frac{2 \log M}{ d \left( \log(1+\alpha) - \frac{\alpha}{1+\alpha} \right)}.
\end{equation}
and 
$x_l(N)$ is the negative solution of 
\begin{equation}
x_l(N) - \log(1+x_l(N)) = \frac{2 \log M}{Nd},  \quad -1 < x_l(N) < 0.
\end{equation}
This nonlinear equation has another solution at $x_u (N) > 0 $.
At the transition point
\begin{equation}
x_l(N_c) = - \frac{\alpha}{1+\alpha}. 
\end{equation}

The free energy difference for an {\it isothermal} volume change
computed from the values of work measured in the 
{\it adiabatic} process is
\begin{equation}
\Delta F = \lim_{M \rightarrow \infty} \mathbb{E}\left[\Delta F_{M}\right] = 
N \frac{d}{2} log(1+\alpha) = N \log\frac{V_0}{V_1}.
\end{equation}
in units of $\beta^{-1} = k_B T$.

The bias in the Jarzynski estimator of the free energy difference is
\begin{equation}
B_{M} =
\left\{
\begin{array}{ll}
0,                                                                      & N \le N_c \\
N \frac{d}{2} \left(\alpha (1+x_l) - log(1+\alpha)\right) + \log M,     & N  > N_c  
\end{array}
\right.
\end{equation}
The maximum value of the bias corresponds to $M = 1$, which implies
$x_l = x_u = 0$, $N_c =0$, and
\begin{equation}
B_{\max} = B_{1} = \left< W \right> - \Delta F  =  
N \frac{d}{2} \left(\alpha - log(1+\alpha)\right).
\end{equation}

Therefore, the bias normalized by its maximum value is
\begin{equation}
{\tilde B}_{M} = \frac{B_{M}}{B_{\max}} 
= 
\left\{
\begin{array}{ll}
 0,               & N \le N_c \\
1 + \frac{ x_l (1 + \alpha ) - \log (1+x_l)}{\alpha - \log(1+\alpha)}, & N  > N_c  
\end{array}
\right.
.
\end{equation}

\begin{figure}
\includegraphics[scale=0.55]{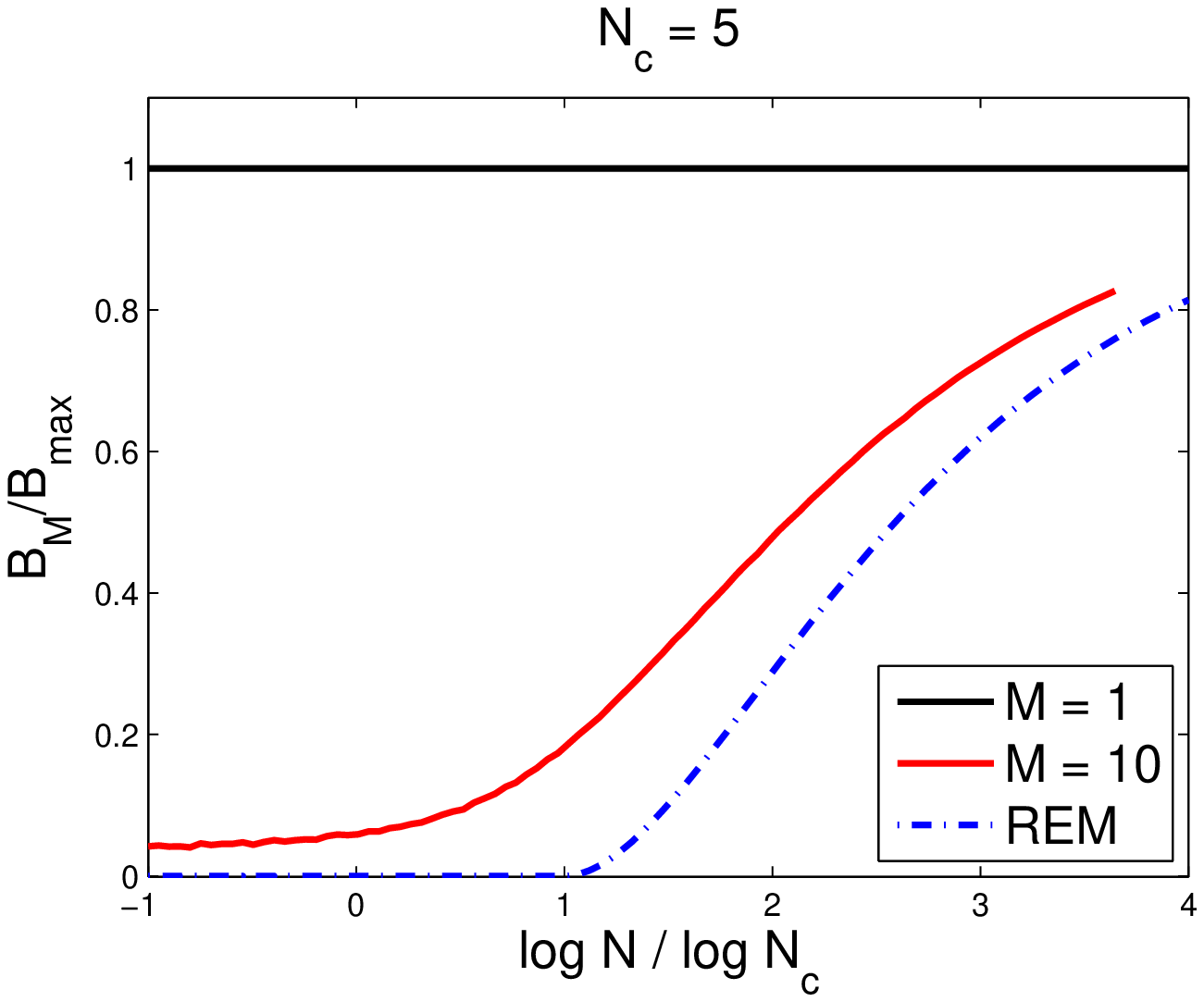} 
\includegraphics[scale=0.55]{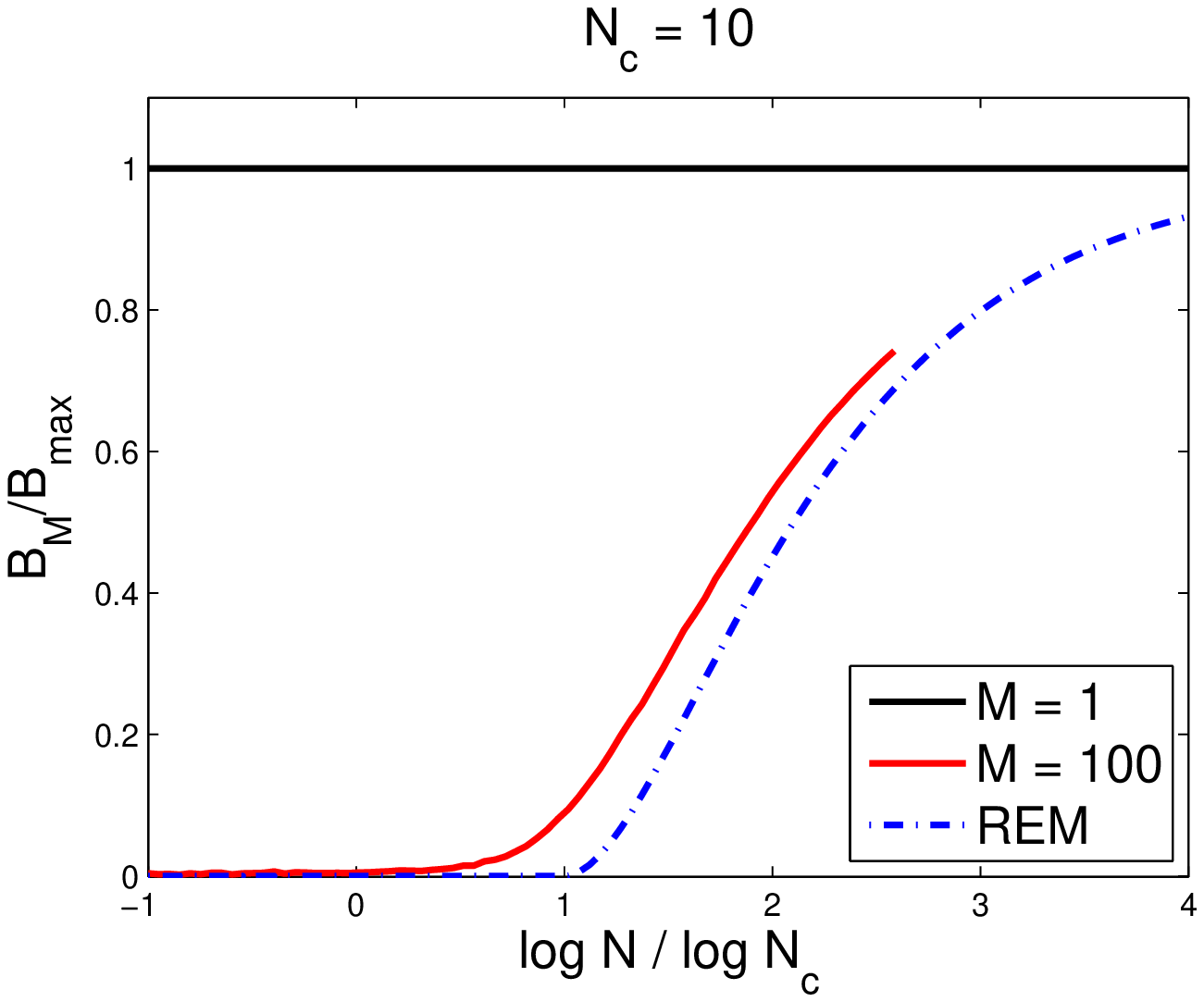}   
\includegraphics[scale=0.55]{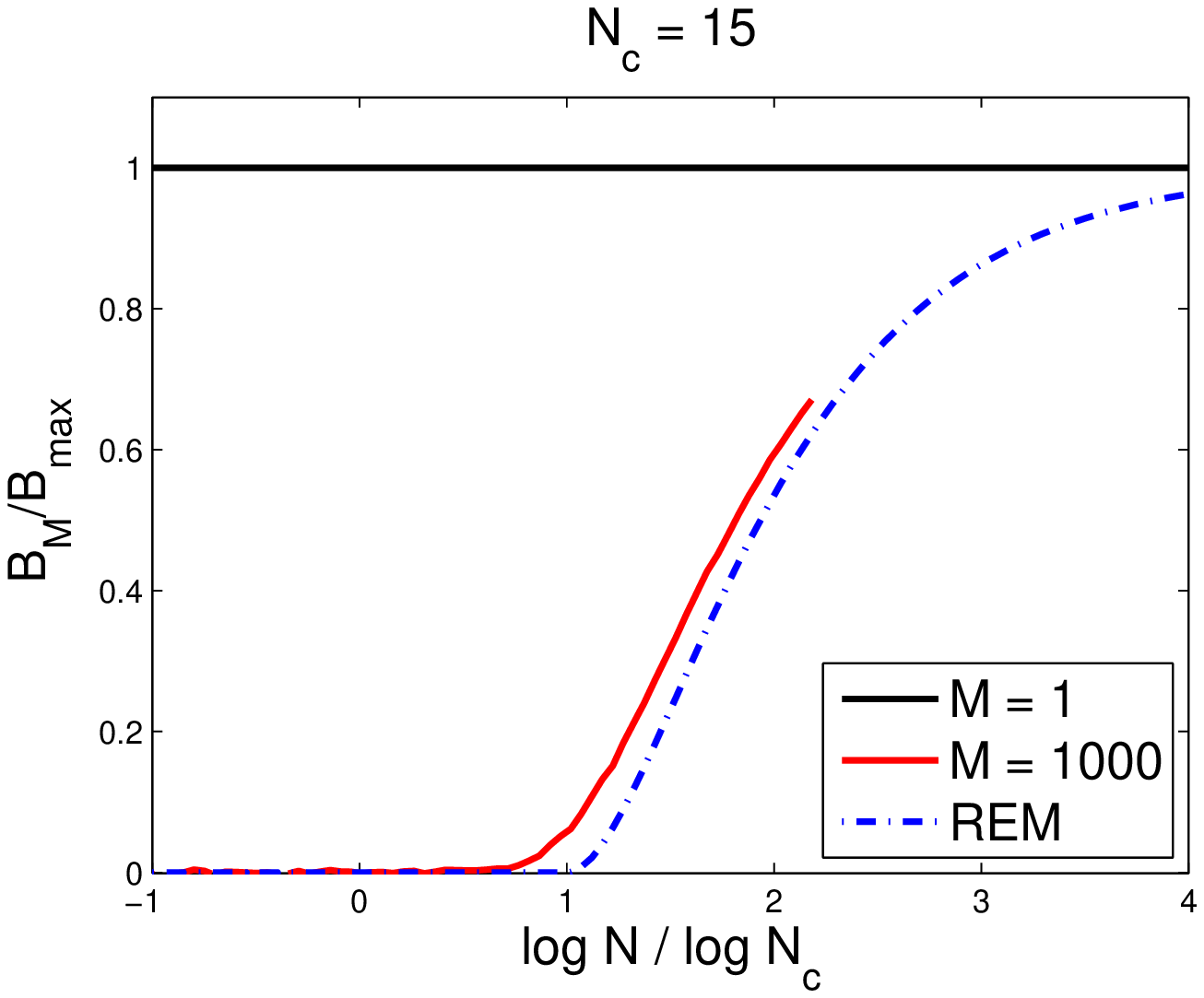} 
\includegraphics[scale=0.55]{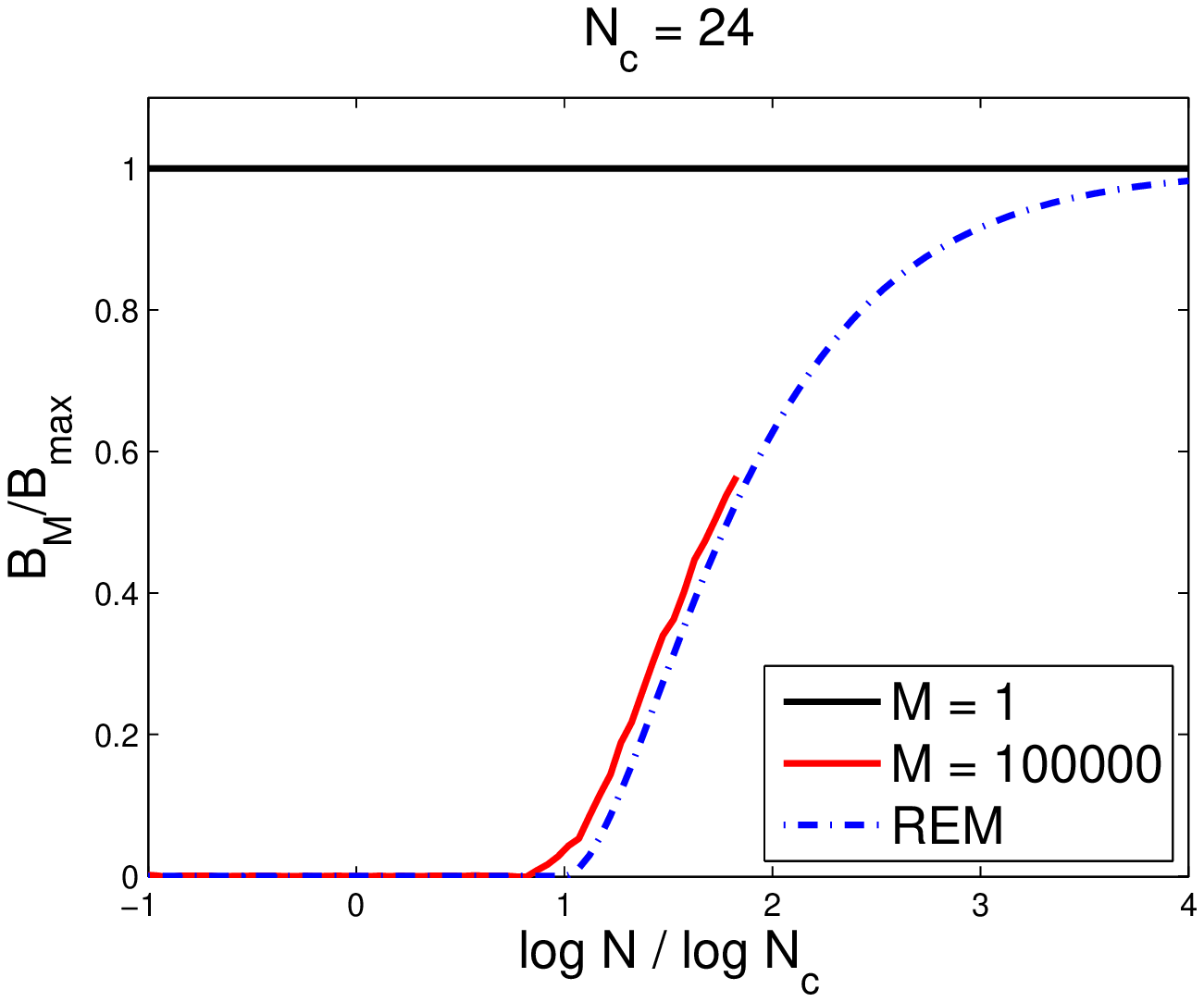} 
\caption{Dependence of the normalized bias on $N$ for 
$M = 10$ (top left), $M = 100$ (top right), $M = 1000$ (bottom left) and
$M = 100,000$ (bottom right).
The dash-dotted curve corresponds to the  
discrete random energy model and gives the limiting behavior as $M \rightarrow \infty$.
}\label{fig:phaseTranstion_Gamma_REM}
\end{figure}

The results obtained in simulations of the adiabatic compression
of a dilute real gas of $N$ particles in  $d=3$ dimensions
from a volume $V_0$ to a volume $V_1  = V_0/4 $ are shown in 
Figure \ref{fig:phaseTranstion_Gamma_REM}. The
plots display the normalized bias as a function of $\log N / \log N_c$
for different values of $M$, the number of measurements.
In contrast with the Gaussian case, the curve for 
the random energy model changes, albeit slightly, 
as a function of $M$.
The remaining curves in the plots correspond to averages over
Monte Carlo simulations. The simulations cannot be
performed for large values of $N$ because of numerical
overflows. For fixed values of $M$, one observes a 
transition from a regime in which the bias is close to zero, 
for small systems, to a region in which the Jarzynski 
estimator is biased, when the system size is above a 
threshold $N_c$. The value of $N_c$  has a logarithmic 
dependence on $M$, which means that the
change in regime occurs for fairly small systems, even 
when the number of experiments is rather large
($N_c \approx 24$ for $M=100,000$).
Beyond this threshold, the bias increases monotonically
and asymptotically approaches 
its maximum. The regime in which the bias is
maximal corresponds to the realm of classical
thermodynamics. Within this regime
the measured work can be identified with
either the average ($(Nd/2) \alpha$) or  
the typical value  ($(Nd/2 -1) \alpha$).
Typical and average values of work, as well 
as the Jarzynski estimator in a finite
number of measurements become indistinguishable
as the number of gas particles increases.
 
The change between these two regimes, which is smooth for 
small $M$, becomes more abrupt as $M$ increases. The
variant of the REM model analyzed in Appendix \ref{app:Gamma_REM}
provides a good qualitative description of
the transition that becomes more accurate as
$M$, $N$ and the volume change become larger.

\section{Discussion and conclusions \label{sec:conclusions}}

There are two paradigms in which statistical mechanics
has successfully bridged mechanical and thermodynamic 
laws in systems composed of large numbers of particles
using a probabilistic description. 
On the one hand, individual macroscopic systems 
can be characterized by the values of thermodynamic quantities
and their fluctuations. 
In these large systems the distribution of each of the 
thermodynamic quantities is sharply peaked around its mode. 
Therefore, the fluctuations are small and need to be probed 
by means of special experimental techniques (e.g. light 
scattering for mass density fluctuations) or in indirect 
measurements (e.g. using fluctuation-dissipation relations 
to analyze the relaxation of systems removed from equilibrium). 
The most probable value obtained in a single experiment agrees 
with the average over several experiments.
On the other hand, in a small system coupled 
to a large number of degrees of
freedom the standard thermodynamic description is 
valid for the system plus environment considered as a whole. 
However, measurements of reduced properties 
associated with a small number of degrees of freedom 
are dominated by fluctuations. 
In consequence, thermodynamic relations cannot be 
used to describe individual measurements. 
Notwithstanding, they can be understood in terms of
averages and fluctuations over repeated measurements. 
The coupling of the small system to the environment provides 
a mechanism for the reduced properties to approach 
their equilibrium values as given by the corresponding
ensemble average \cite{prigogine+henin_1960_general,henin++_1960_general,ronis+oppenheim_1977_nonlinear,oppenheim_1989_nonlinear}.
Each measurement can be understood as sampling the initial 
state from the equilibrium distribution
for the appropriate constraints (e.g. temperature, 
if the environment acts as a thermal reservoir).   
For standard thermodynamic quantities, these measurements
over ensembles of small systems yield typical and average 
values that are close to each other. 
In contrast, typical values obtained 
in individual measurements of work in nonequilibrium 
processes do not fulfill the Jarzynski equality.
In this equality, the average of the 
exponential of minus the work is computed by taking 
the mean of repeated independent measurements in
systems prepared at equilibrium. By the law of
large numbers, this Monte Carlo estimate converges
to the expected value in the limit of an infinite number of
measurements. 
However, the sum is dominated by rare events that need 
to be sufficiently well sampled so that the Monte Carlo 
estimate is close to the asymptotic result.  
If the number of samples is below a certain
threshold, the estimator is biased. It  becomes
unbiased only for sufficiently large sample sizes.
In the appropriate scaling limit, the change of behavior 
corresponds to a phase transition in 
variants of the random energy model, a simplified
model for spin glasses.

This phenomenon can be analyzed from an alternative point
of view. Assuming that the resources available are limited
and that the number of measurements $M$ is fixed, 
the change of behavior in the estimator
appears as the size of the system is
increased. For small systems, the thermodynamic 
description needs to be understood in terms of ensemble
averages, not of individual measurements.
In this regime, the Jarzynski  estimator is unbiased. It
is equal to the exponential of minus the free energy 
difference in an isothermal evolution between
the initial and the final values of the work parameter, 
independently of other characteristics of the actual 
nonequilibrium process that takes place in the system.
In particular, it does not depend on the final state
of the system after the external manipulation is 
completed. The lack of sensitivity to the details of the 
manipulation is a striking reflection at the 
microscopic level of the Hamiltonian evolution of the 
system as a whole (including the degrees of freedom 
of the environment) as demonstrated in \cite{jarzynski_2004_nonequilibrium}. 
The mechanical character
of the equality is also highlighted by the fact
that the largest contributions to the average 
correspond to initial configurations of the system that 
are typical of the reverse process \cite{jarzynski_2006_rare}. 
Since these configurations need to be sufficiently 
well sampled, this implies that, under these conditions, 
the arrow of time is only tenuously defined. 

For large systems and fixed $M$, 
the Jarzynski estimator becomes biased. 
In fact, as the size of the system
approaches infinity, the average of the exponential of
minus the work values obtained in the nonequilibrium 
process approaches the exponential of minus the average
work. Since the typical and the average are close to 
each other, a single experiment and an average over
a number of experiments of the order of $M$ yield
results that are equivalent within measurement errors.
Furthermore, this average depends on the details
of the external manipulation. These features
are in agreement with the standard thermodynamic 
description of an individual macroscopic system:
the dominant configurations, 
which are rare for the initial state but typical 
of the final equilibrium state, are never sampled. 
The sample mean is dominated by typical configurations,
which are characteristic of the forward process.
Therefore, in this regime, the arrow of time becomes well defined.

The necessary condition to observe a transition in the Jarzynski 
estimator is that $<W>$, the average work in the nonequilibrium process,
be different from $ \Delta F$, the isothermal free-energy change. 
If the non-equilibrium process is adiabatic, these quantities
are different even when the external manipulation of the system is slow. 
This can be seen in section \ref{subsec:adiabatic}, 
which analyzes a quasi-static adiabatic volume change in 
a dilute gas. Another example is the adiabatic expansion of an ideal 
gas against a piston: The maximum bias $ B_{max} = <W> - \Delta F$
is different from zero even when the velocity of the piston approaches 0 
(see Figure 4 in \cite{palmieri+ronis_2007_jarzynski}). 
The situation is different for isothermal processes. In this case 
$<W>$ approaches $\Delta F$ in the limit of quasi-static driving. 
Therefore, in an isothermal process, one should expect a transition 
from a fast driving regime, in which the Jarzynski estimator 
of the free energy difference is biased, to a slow driving regime,
in which the Jarzynski estimator is unbiased. Isothermal processes 
are currently under analysis.

The conclusions of this study are, of course, not new. The emergence
of irreversibility and of a thermodynamic description in 
systems with many degrees of freedom is one of the central 
results of statistical mechanics 
\cite{boltzmann_1872_further,gibbs_1902_elementary,khinchin_1949_mathematical,zwanzig_2001_nonequilibrium,castiglione++_2008_chaos}.
The analysis carried out shows how, in
the context of the Jarzynski equality, the emergence
of this macroscopic picture can be understood 
in terms of a phase transition that appears
in the proper scaling limit and when the measurement 
resources are limited as the size of the system increases.

\appendix
\section{The random energy model} \label{app:REM}
The random energy model (REM) was introduced in 
\cite{derrida_1980_random,derrida_1981_random} as a simplified model
for spin glasses that captures many salient properties of these types 
of disordered systems. In the random energy model, the
system has $M = 2^K$ energy levels, $\left\{ E_i \right\}_{i=1}^{M}$. 
These energy levels are independent identically distributed 
random variables sampled from a normal distribution
\begin{equation}
p(E) = \frac{1}{\sqrt{\pi K}} e^{-E^2/K}.
\end{equation}
The canonical partition function for a particular system (i.e.
for a particular  realization of the $M$ energy levels) 
at temperature $\beta^{-1}$ is
\begin{equation}
Z_M(\beta) \equiv \sum_{i = 1}^{2^K} e^{-\beta E_i}.
\end{equation}

In the limit of large $K \rightarrow \infty$, the system 
undergoes a second order phase transition
\begin{equation}
\lim_{K \rightarrow \infty}
\frac{1}{K}\mathbb{E} \left[\log Z_M(\beta) \right] =
\left\{
\begin{array}{ll}
 \frac{\beta^2}{4} + \log 2,   & \beta \le \beta_c \\
 \beta \sqrt{\log 2},          & \beta > \beta_c 
\end{array}
\right.
,
\end{equation}
where $\beta_c = 2 \sqrt{\log 2}$.

To make the connection between the Gaussian REM and the empirical estimation of the 
free energy in a process in which the work distribution is Gaussian,
$W_m  \sim N\left(\left< W\right>, \sigma^2 \right)$
we make the change of variables
\begin{eqnarray}
W_m  & = &  \left< W \right> + \sqrt{\frac{2 \log 2}{\log M}} \sigma E_m \nonumber \\ 
E_m & \sim & N\left(0, \frac{\log M}{2 \log 2}  \right) 
, 
\end{eqnarray}
where we have used the fact that $K = \log M / \log 2$.
In terms of these new variables 
\begin{eqnarray}
\left< e^{-W}\right>_M = \frac{1}{M} \sum_{m=1}^M e^{-W_m} = 
\frac{1}{M} e^{- \left< W \right>} \sum_{m=1}^M e^{- \sqrt{\frac{2 \log 2}{\log M}} \sigma E_m}. 
\end{eqnarray}  
Identifying $ \sqrt{\frac{2 \log 2}{\log M}} \sigma$ with $\beta$ in the REM,
we obtain 
\begin{eqnarray}
\mathbb{E}\left[\Delta F_{M}\right] & = &  
- \mathbb{E} \left[ \log  \left< e^{-W}\right>_M  \right] 
=  \left< W \right> + \log M - 
\mathbb{E} \left[ \log \left( \sum_{m=1}^M 
e^{-\sqrt{\frac{2 \log 2}{\log M}} \sigma E_m} \right) \right]  \nonumber \\
& = & \left\{ 
\begin{array}{ll}
\left< W \right> - \frac{1}{2} \sigma^2, 
&   M \ge \exp \left\{ \frac{1}{2} \sigma^2 \right\} \\
\left< W \right> - \sqrt{2} \sigma \sqrt{\log M} + \log M,  
 & M < \exp \left\{ \frac{1}{2} \sigma^2 \right\}
\end{array}
\right.
\end{eqnarray}
for $M \rightarrow \infty$ and $ \sigma^2 \rightarrow \infty$ with $ \log M / \sigma^2 \rightarrow \hbox{constant}$.
 
\section{The random energy model with energy levels that follow a gamma distribution}
\label{app:Gamma_REM}
Consider a system with $M$ random energy levels, 
$\left\{ E_i \right\}_{i=1}^{M}$, 
independently sampled from a gamma distribution
\begin{equation} \label{eq:gammaDistributionZeroMean}
p(E) =  \frac{\left( E + K \right)^{K-1}}{\Gamma(K)}  e^{- \left(E + K\right)}, 
\quad  E \in [-K,\infty).
\end{equation}

Define the parameter $\xi = \frac{\log M}{K \log 2}$. 
In terms of this parameter $M = 2^{\xi K}$.
It is possible to derive an expression for the entropy in the 
limit $K \rightarrow \infty$ by analyzing the behavior of
${\cal N}(\epsilon,\epsilon + \delta)$, the number of energy levels 
in an interval $ {\cal I}(\epsilon;\delta)   = 
\left[K \epsilon, K (\epsilon + \delta) \right]$, with $\epsilon \ge - 1 $, 
$\delta >0 $ \cite{mezard+montanari_2009_information}.
Since this count depends on the realization
of the system, ${\cal N}(\epsilon,\epsilon + \delta)$
is a binomial random variable whose first two moments
are
\begin{eqnarray} 
\mathbb{E}\left[ {\cal N}(\epsilon,\epsilon + \delta) \right] 
& = & 2^{\xi K} {\cal P}_{\cal I}(\epsilon; \delta) \\
\hbox{Var}\left[ {\cal N}(\epsilon,\epsilon + \delta) \right] 
& = & 2^{\xi K} {\cal P}_{\cal I}(\epsilon; \delta) \left(1-{\cal P }_{\cal I}(\epsilon; \delta) \right),
\end{eqnarray} 
where
\begin{equation}
{\cal P}_{\cal I}(\epsilon; \delta) =  
\frac{K^K}{\Gamma(K)} \int_{\epsilon}^{\epsilon + \delta}  (x+1)^{K-1} e^{-K(x+1)} dx
\end{equation}
is the probability of an individual energy level to be in the 
interval $ {\cal I}(\epsilon;\delta) $.
As $K \rightarrow \infty$ these moments can be approximated to leading
exponential order as
\begin{eqnarray}
\mathbb{E}\left[ {\cal N}(\epsilon,\epsilon + \delta) \right] 
& \doteq & \exp \left\{ K \max_{[\epsilon, \epsilon+\delta]} s_a(x) \right\} \\
\frac{\hbox{Var}\left[ {\cal N}([\epsilon,\epsilon + \delta]) \right]}{\left[\mathbb{E}\left[{\cal N}(\epsilon,\epsilon + \delta) \right]\right]^2} 
& \doteq & \exp \left\{ -K \max_{[\epsilon, \epsilon+\delta]} s_a(x) \right\} 
\end{eqnarray}
with 
\begin{equation}
s_a(x) = \xi \log 2 + \log(1+x) - x.
\end{equation}
Note that $s_a(x) \ge 0$ for $x_l \le x \le x_u$,
where $ -1 \le x_l \le 0 \le x_u $ fulfill 
$s_a(x_l) = s_a(x_u) = 0$.
The limiting behavior of this equation when $\xi \rightarrow 0$ is 
\begin{equation}
s_a(x) = \xi \log 2 - x^2 /2  \quad x_l \le x \le x_u,
\end{equation}
where $x_l  \approx  - \sqrt{2 \xi \log 2 } $ and
$ x_u \approx  \sqrt{2  \xi \log 2} $. In this limit 
the results are similar to the Gaussian REM.
In the opposite limit, when $\xi \rightarrow \infty$ is 
\begin{equation}
s_a(x) = \xi \log 2 + log(1+x),  \quad x  \approx x_l, 
\end{equation}
where $x_l \approx -1 + 2^{-\xi}$.

The entropy function is defined as 
\begin{equation}
s(\epsilon) = \left\{ 
\begin{array}{ll}
s_a(x) = \xi \log 2 + \log(1+x) - x, 
& \quad x_l \le x \le x_u \\ 
-\infty, & \quad  x < x_l, x > x_u.
\end{array}
\right.
\end{equation}
It can be shown that for any pair $\epsilon$, $\delta$, with probability one,
\begin{equation}
\lim_{K \rightarrow \infty} \frac{1}{K} \log {\cal N}(\epsilon,\epsilon + \delta) = \sup_{[\epsilon,\epsilon+\delta]} s(x). 
\end{equation}

\begin{figure}
\includegraphics[scale=1]{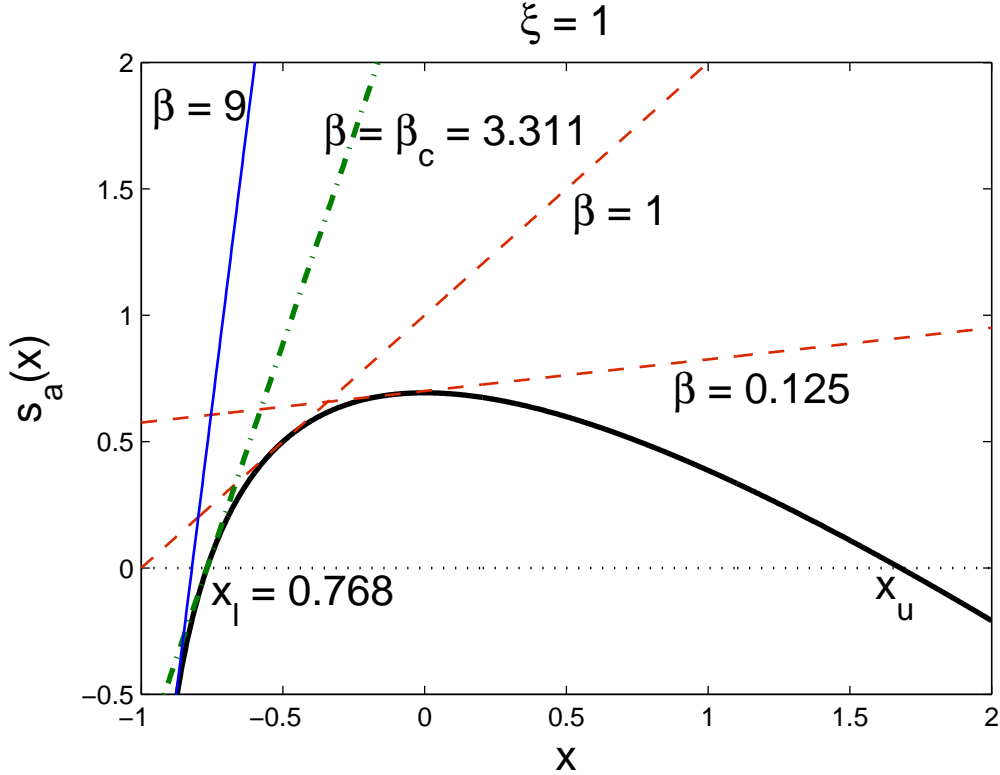}
\caption{Transition between the low and high temperature regimes
in the random energy model with energy levels sampled from  a Gamma 
distribution (see text for details).}.
\label{fig:graphicalPhaseTransition_Gamma_REM}
\end{figure}

The canonical partition function for a particular
realization of the $M$ energy levels 
at temperature $\beta^{-1}$ is
\begin{equation}
Z_M(\beta) = \sum_{i = 1}^{2^{\xi K}} e^{-\beta E_i}.
\end{equation}
In the limit $K \rightarrow \infty$,
\begin{equation}
Z_M(\beta) \doteq \int_{x_l}^{x_u} 
\exp\left[K \left(s_a(x) - \beta x\right) \right] dx
\doteq \exp\left[K \max_{x \in [x_l,x_u]}
\left(s_a(x) - \beta x \right) \right] 
\end{equation}
to exponential accuracy. Depending on the temperature, 
the maximum is either in the interval $(x_l,0)$ (high temperature)
or at $x_l$ (low temperature)
\begin{equation} \label{eq:max_REM_Gamma}
{\arg\max}_{x \in [x_l,x_u]} \left(s_a(x) - \beta x \right) = 
\left\{
\begin{array}{ll}
-\frac{\beta}{1+\beta} , & \beta \le \beta_c\\
x_l, & \beta > \beta_c
\end{array}
\right.
,
\end{equation}
where $\beta_c = -x_l/(1+x_l)$.
A graphical construction that illustrates this transition is presented
in figure (\ref{fig:graphicalPhaseTransition_Gamma_REM}) for $\xi = 1$.
The curves displayed are $s_a(x)$ and  straight lines with slope $\beta$  
that are tangent to $s_a(x)$ at the points that are local maxima
of $s_a(x)- \beta x$. For high temperatures ($\beta \le \beta_c$) the
local maximum is within the interval $[x_l,x_u]$ and is therefore
the solution of  (\ref{eq:max_REM_Gamma}). At low temperatures 
($ \beta > \beta_c$), the solution to (\ref{eq:max_REM_Gamma}) is $x_l$,
which, in this regime, is a global maximum, but not a local one.  
Using these relations it can be shown that, in this limit, 
the system undergoes a second order phase transition
\begin{equation}
\lim_{K\rightarrow \infty} \frac{1}{K}\mathbb{E} \left[\log Z_M(\beta) \right] =
\left\{
\begin{array}{ll}
\xi \log 2 +  \beta - log(1+\beta),     & \beta \le \beta_c \\
   - \beta x_l  = \beta \beta_c / (1+\beta_c),                 & \beta > \beta_c 
\end{array}
\right.
.
\end{equation}
Continuity at $\beta_c$ implies that
\begin{equation}
log(1+\beta_c) - \frac{\beta_c}{1+\beta_c} = \xi \log 2.  
\end{equation}

Consider now a sample of $M = 2^{\xi K}$ work values $\left\{ W_m \right\}_{m=1}^M$ 
from the distribution
\begin{equation}
p(W) =  \frac{1}{\left|\alpha\right|\Gamma(K)}
\left(\frac{W}{\alpha}\right)^{K-1}  e^{- W / \alpha} \theta(\alpha W),
\end{equation}
with $K = Nd/2$.
The connection with the variant of REM analyzed in this section is 
achieved by making the change of variable 
\begin{eqnarray}
W_m  & = &  \left< W \right> + \alpha  E_m \quad m = 1,\ldots,M,
\end{eqnarray}where $ \left< W \right> = N \frac{d}{2} \alpha$
is the average work and $E_m$ are iidrvs sampled from the distribution
(\ref{eq:gammaDistributionZeroMean}). 
In terms of these new variables 
\begin{eqnarray}
\left< e^{-W}\right>_M = \frac{1}{M} \sum_{m=1}^M e^{-W_m} = 
\frac{1}{M} e^{- \left< W \right>} \sum_{m=1}^M e^{- \alpha E_m}. 
\end{eqnarray}  
Identifying $ \alpha $ with $\beta$ in the random energy model, 
we conclude that there is an abrupt change of behavior of
the Jarzynski estimator of the free energy differences
for $ M \rightarrow \infty$, $ N \rightarrow \infty$ with
$\log M / N \rightarrow \hbox{constant}$,
as a function of $\xi = \frac{2 \log M}{N d \log 2}$ 
\begin{eqnarray}
\mathbb{E}\left[\Delta F_{M}\right] & = & -\mathbb{E} \left[\log \left< e^{-W}\right>_M \right] =
\left< W \right> + \log M - \mathbb{E} \left[\log \sum_{m=1}^M 
e^{- \alpha E_m} \right]  \nonumber \\
& = & 
\left\{
\begin{array}{ll}
 N \frac{d}{2} log(1+\alpha),                     & \xi \ge \xi_c \\
 N \frac{d}{2} \alpha (1+x_l(\xi)) + \log M,   & \xi  <  \xi_c  
\end{array}
\right.
,
\end{eqnarray}
where   
\begin{equation}
\xi_c = \frac{ \log(1+\alpha) - \frac{\alpha}{1+\alpha}}{\log 2}.
\end{equation}
and $x_l(\xi)$ is the negative solution of the nonlinear equation
\begin{equation}
x_l(\xi) - \log(1+x_l(\xi)) = \xi \log 2, \quad -1 < x_l(\xi) < 0.
\end{equation}
At the transition point $ \xi = \xi_c$
\begin{equation}
x_l(\xi_c) = - \frac{\alpha}{1+\alpha}. 
\end{equation}
For a fixed number of particles $N$, the transition takes place 
when the number of measurements is above the threshold
\begin{equation}
M_c = \exp\left[ N \frac{d}{2} \left( \log(1+\alpha) - \frac{\alpha}{1+\alpha} \right) \right].
\end{equation}

Alternatively, for fixed number of experiments $M$, the
transition occurs when the number of particles in the
gas is below 
\begin{equation}
N_c = \frac{2 \log M}{ d \left( \log(1+\alpha) - \frac{\alpha}{1+\alpha} \right)}.
\end{equation}

\section{The discrete random energy model} \label{app:DREM}
Consider a system  with $M = 2^K$ energy levels, 
$\left\{ E_i \right\}_{i=1}^{M}$. 
These energy levels are independent identically distributed 
random variables sampled from a binomial distribution
\begin{equation}
p(E) = \frac{1}{2^N} \binom{N}{\frac{1}{2}N + E}, \quad 
E = -\frac{N}{2},-\frac{N}{2} +1,\ldots, \frac{N}{2}.
\end{equation}
The canonical partition function at temperature $\beta^{-1}$ is
\begin{equation}
Z_M(\beta) = \sum_{i=1}^{2^K} e^{-\beta E_i}.
\end{equation}

In the limit 
\begin{equation}
M \rightarrow \infty \quad N \rightarrow \infty, 
\quad \gamma = \frac{K}{N} =\frac{\log M}{N \log 2}  \rightarrow  \hbox{constant}
\end{equation} 
the system undergoes a second order phase transition \cite{ogure+kabashima_2009_analyticity_1}
\begin{equation}
\mathbb{E} \left[\log Z_M(\beta) \right] =
\mathbb{E} \left[\log \left( \sum_{i = 1}^{2^K} e^{-\beta E_i} \right) \right] = 
\left\{
\begin{array}{ll}
K  \log 2 +  N \log \cosh \frac{\beta}{2},   & \beta \le \beta_c \\
\frac{1}{2}\beta N \tanh \frac{\beta_c}{2},    & \beta > \beta_c 
\end{array}
\right.
\end{equation}
where 
\begin{equation}
\beta_c =  
\left\{
\begin{array}{ll}
\infty,        & \gamma \ge 1 \\
\log \left[1- h_2^{-1}(1-\gamma) \right] 
- \log \left[ h_2^{-1}(1-\gamma) \right], \quad & \gamma  < 1  
\end{array}
\right.  
\end{equation}
and the function $h_2^{-1}(y) \in [0, 1/2]$ is the inverse of the binary entropy
\begin{equation}
h_2(x) =  -x \log_2 x -(1-x) \log_2(1-x).
\end{equation}

To make the connection with the ideal gas compression experiment
in the calculation of 
\begin{equation}
\mathbb{E}\left[\Delta F_{M}\right] = - \mathbb{E} \left[\log \left< e^{-W} \right>_M \right] = 
- \mathbb{E} \left[ \log  \left(\frac{1}{M} \sum_{m=1}^M e^{-n_m \epsilon}\right) \right],
\end{equation}
where $n_m \in \left\{0,1, \ldots N \right\}$ 
follows a binomial distribution of parameters
$N$, $p(0) = 1/2$,
we make the change of variable
\begin{equation}
n_m = \frac{1}{2} N + E_m; \quad E_m \in 
\left\{-\frac{N}{2},-\frac{N}{2}+1, \ldots \frac{N}{2} \right\}.
\end{equation}
Using these new random variables
\begin{eqnarray}
\mathbb{E}\left[\Delta F_{M}\right] & = & 
 - \mathbb{E} \left[\log \left( e^{- \frac{1}{2} N \epsilon}
\frac{1}{M} \sum_{m = 1}^M  e^{- E_m \epsilon} \right)\right] \nonumber \\
& = & 
N \frac{\epsilon}{2}   +  \log M -  
\mathbb{E} \left[\log \left(\sum_{m = 1}^M  e^{- E_m \epsilon} \right)\right].
\end{eqnarray}

Finally, identifying $\epsilon$ with $\beta$ in the DREM, 
and $\epsilon_c = \beta_c$
\begin{equation}
\mathbb{E}\left[\Delta F_{M}\right] 
= 
\left\{
\begin{array}{ll}
N \left[\frac{\epsilon}{2}  -   \log \cosh \frac{\epsilon}{2} \right] 
= N \log \frac{2}{1 + e^{-\epsilon}},   
 & \epsilon \le \epsilon_c \\
N \left[\gamma \log 2 + 
 \frac{\epsilon}{2}  \left(1 -   \tanh \frac{\epsilon_c}{2} \right)
\right], & \epsilon > \epsilon_c
\end{array}
\right. 
,
\end{equation}
where we have used that fact that $K = \log M / \log 2$ and $ \gamma = K/N = \frac{\log M}{N\log 2}$.

\begin{acknowledgments}
The authors thank Eric Zimanyi for insightful discussions. A.S. acknowledges partial financial support from the Spanish \emph{Direcci\'on General de Investigaci\'on},  project TIN2010-21575-C02-02.
\end{acknowledgments}



%
%
%

\end{document}